\documentstyle[mnextra]{mn}

\seceqnum	

\input{epsf}

\newdimen\inmarginw \inmarginw=1.5 truecm 
\def\strutdepth{\dp\strutbox}
\def\inmargin#1{\strut\vadjust{\kern-\strutdepth{\vtop to
\strutdepth{\baselineskip%
\strutdepth\vss\llap{\hbox to \inmarginw{#1\hss}\space}\null}}}}

\def\v178{V_{178}}
\def\m178{M_{178}}
\def\r178{r_{178}}
\def\mg{M_{\rm group}}
\def\mstar{M_{*}}
\def\rd{s}
\def\vc{V_{\rm c}}

\def\etal{et al.\ }
\def\nfw{NFW\ }
\def\s1d{\sigma_{\rm v}^{\rm 1D}}
\def\rt2{2^{1/2}}
\def\eg{e.g.\ }
\def\ie{i.e.\ }
\def\an{a_{\rm N}}
\def\ah{a_{\rm H}}
\def\bn{b_{\rm N}}
\def\bh{b_{\rm H}}
\def\msun{M_{\odot}}
\def\lta{\mathrel{\hbox{\rlap{\hbox{\lower4pt\hbox{$\sim$}}}\hbox{$<$}}}}
\def\gta{\mathrel{\hbox{\rlap{\hbox{\lower4pt\hbox{$\sim$}}}\hbox{$>$}}}}
\def\lsim{\mathrel{\hbox{\rlap{\hbox{\lower4pt\hbox{$\sim$}}}\hbox{$<$}}}}
\def\gsim{\mathrel{\hbox{\rlap{\hbox{\lower4pt\hbox{$\sim$}}}\hbox{$>$}}}}
\begin{document}
\title[The Structure of Haloes]{The Structure of Dark Matter Haloes in
Hierarchical Clustering Models}

\author[S. Cole and C. Lacey]{Shaun Cole$^{1,4}$ and Cedric
Lacey$^{2,3,5}$\\
$^1$Department of Physics, University of Durham, Science
Laboratories, South Rd, Durham DH1 3LE\\
$^2$Physics Department, University of Oxford, Keble Rd, Oxford OX1 3RH\\
$^3$Theoretical Astrophysics Center, Blegdamsvej 17,
DK-2100 Copenhagen O, Denmark.\\
$^4$Shaun.Cole@durham.ac.uk\\
$^5$lacey@tac.dk}

\maketitle

\begin{abstract}
 We use a set of large cosmological N-body simulations to study the
internal structure of dark matter haloes which form in scale-free
hierarchical clustering models (initial power spectra $P(k) \propto
k^n$ with $n=0$,$-1$ and~$-2$) in an $\Omega=1$ universe. We find that
the radius $r_{178}$ in a halo corresponding to a mean interior
overdensity of 178 accurately delineates the quasi-static halo
interior from the surrounding infalling material, in agreement with
the simple spherical collapse model. The interior velocity dispersion
correlates with mass, again in good agreement with the spherical
collapse model. Interior to the virial radius $\r178$, the spherically
averaged density, circular velocity and velocity dispersion profiles
are well fit by a simple two-parameter analytical model proposed by
Navarro \etal (1995a). This model has $\rho \propto r^{-1}$ at small
radii, steepening to $\rho \propto r^{-3}$ at large radii, and fits our
haloes to the resolution limit of the simulations.  The two model
parameters, scalelength and mass, are tightly correlated. Lower mass
haloes are more centrally concentrated, and so have scalelengths which
are a smaller fraction of their virial radius than those of their
higher mass counterparts. This reflects the earlier formation times of
low mass haloes.  The haloes are moderately aspherical, with typical
axial ratios $1:0.8:0.65$ at their virial radii, becoming gradually
more spherical towards their centres. The haloes are generically
triaxial, but with a slight preference for prolate over oblate
configurations, at least for $n=-1$ and~$0$. These shapes are
maintained by an anisotropic velocity dispersion tensor. The median
value of the spin parameter is $\lambda
\approx 0.04$, with a weak trend for lower $\lambda$ at higher halo
mass. We also investigate how the halo properties depend on the
algorithm used to identify them in the simulations, using both
friends-of-friends and spherical overdensity methods. We find that for
groups selected at mean overdensities $\sim 100-400$ by either method,
the properties are insensitive to how the haloes are selected, if the
halo centre is taken as the position of the most bound particle.

\end{abstract}

\begin{keywords}
cosmology: theory -- dark matter
\end{keywords}

\section{Introduction}

The structure of dark matter haloes is of fundamental importance to
understanding the formation and evolution of galaxies and galaxy
clusters. In hierarchical clustering models, galaxies form by gas
cooling and condensing in dark matter haloes, and clusters form by the
gravitational aggregation of individual galaxies and galaxy groups.
This evolution is driven by gravitational instability via accretion
and merging. Many studies have been made of the properties of haloes
formed by dissipationless gravitational collapse using N-body
simulations, which are ideally suited to investigate this inherently
non-linear problem. Continuing advances in computer technology and
codes have enabled ever larger simulations to be performed, allowing
halo structure to be examined in ever increasing detail.

Pioneering N-body studies of the structure of haloes formed by
hierarchical clustering in an expanding universe were carried out by
Frenk \etal (1985) and Quinn, Salmon \& Zurek (1986). These
concentrated on halo density profiles, and found approximately flat
rotation curves for $\Omega=1$ cold dark matter (CDM) models,
corresponding to density varying with radius roughly as $\rho\propto
r^{-2}$. Frenk \etal (1988) made a detailed study of haloes in CDM
models in flat, open and closed universes, using a P$^3$M code with
$32^3$ particles, calculating rotation curves, shapes and angular
momenta of haloes. Efstathiou \etal (1988) made a similar study of
haloes in self-similar clustering models with $\Omega=1$ and
scale-free initial conditions, meaning initial power spectra of the
form $P(k)\propto k^n$, with $n=1,0,-1$ and $-2$. These studies, and
those of Quinn, Salmon \& Zurek (1986) and Zurek, Quinn \& Salmon
(1988), which used a combination of PM and PP methods and comparable
numbers of particles, found that the slopes of halo density profiles
varied both with spectral index $n$ and with $\Omega$.

More recently, there have been studies with higher resolution and/or
more particles. Warren \etal (1992) simulated $\Omega=1$ scale-free
models using a tree-code and $\sim 10^6$ particles. They concentrated
on the radial variation of shapes and the distribution of angular
momentum in their better-resolved haloes. Crone, Evrard \& Richstone
(1994) used $64^3$ particle $P^3M$ simulations to study the density
profiles of haloes in a variety of cosmological models for scale-free
initial conditions. They fit these density profiles to
power-laws. However, they only examined the most massive haloes in
each simulation. Navarro, Frenk \& White (1995b) used a hierarchical
tree-code to simulate the formation of a small number of haloes of
very different masses in the CDM model at high resolution, again to
study the density profiles. Dubinski \& Carlberg (1991) also simulated
individual haloes at high resolution, but with a much cruder treatment
of the tidal effects of material outside the halo.

In this paper, we analyse the structure of haloes formed in
self-similar clustering models, with $\Omega=1$ and spectral indices
$n=-2,-1,0$. Scale free initial conditions are both conceptually
simple, and, over limited mass ranges, provide useful approximations
to physical models (such as CDM) whose power spectra have a slope
which varies slowly with mass.  For example, in the standard CDM
model, the effective slope is $n\approx -2$ on the scale of galaxy
haloes ($M\sim 10^{12}\msun$) and $n\approx -1$ on the scale of galaxy
clusters ($M\sim 10^{15}\msun$). Such models also have the great
advantage that their self-similar scaling properties can be used to
distinguish physical effects from artifical features introduced by the
limitations of the numerical simulations. Our simulations used the
P$^3$M code of Efstathiou \etal (1985), with $128^3\approx 2\times
10^6$ particles. Unlike Warren \etal (1992), we employ periodic
boundary conditions. As in the previous studies, we pay particular
attention to halo density profiles and rotation curves, but we also
study the radial variation of other dynamical properties, the
departures from spherical symmetry and the dependence of these
properties on initial conditions. We also investigate how these
properties depend on halo mass.

We are interested in haloes as objects in approximate dynamical (or
virial) equilibrium. An important issue not addressed in detail in
previous studies is where one should draw the boundary of the
virialized region of the halo, and what group-finding method works
best for partitioning a simulation into virialized objects. Previous
studies have each just used a single method of identifying the haloes
(although several methods have been tried), and presented results for
that one method, without investigating how the halo properties might
depend on the method used. In the present paper, we consider several
possible criteria for deciding where the boundary of the virialized
region is. We also compare results obtained using two different
group-finding methods, friends-of-friends (Davis \etal 1985) and the
spherical overdensity method of Lacey \& Cole (1994), for a range of
overdensities.

The plan of the paper is as follows: In Section~\ref{sec:models} we
define virial mass and length scales for haloes, and compare three
analytical halo models which we later use to fit our numerical
results. Section~\ref{sec:sims} describes the N-body simulations, and
how we identify the groups within them and construct radial profiles
of halo properties. In Section~\ref{sec:bulk}, we examine the scalings
of bulk properties of the haloes with mass, and make a critical
comparison of the different group finding algorithms.  In
Section~\ref{sec:sphere}, we examine in detail the spherically
averaged halo profiles, which we compare with the analytical models of
Section~\ref{sec:models}.  Section~\ref{sec:lambdas} discusses the
angular momenta of haloes.  Departures from spherical symmetry are
investigated in Section~\ref{sec:asymm}. Our understanding of these
results and their implications are discussed in
Section~\ref{sec:discuss}, and we conclude in Section~\ref{sec:conc}.

\section{Halo Models}\label{sec:models}

The analytical model most often used to describe the formation of
dark matter haloes is the idealised collapse of a uniform, spherically
symmetric overdense region.  For $\Omega=1$, collapse to a singularity
occurs when linear theory would predict an overdensity of $\delta_{\rm
c} = 3/20 \, (12 \pi)^{2/3} \approx 1.69$. If one argues
that realistic amounts of substructure present before the collapse
will lead to violent relaxation, then the virial theorem and energy
conservation imply that after virialisation the object will have a
density of $18 \pi^2 \approx 178$ times the background density $\bar
\rho$, if the final equilibrium state is also a uniform sphere. Real
dark matter haloes are not uniform spheres, but we will refer to the
radius $r_{178}$ around the halo centre within which the mean density
is $178\bar \rho$ as the ``virial radius''.  This terminology will be
justified by the results presented in Section~\ref{sec:sphere}, where
we will see that the radius $\r178$ approximately demarcates the inner
regions of haloes at $r\lta \r178$ which are in approximate dynamical
equlibrium from the outer regions at $r\gta \r178$ which are still
infalling. The corresponding ``virial mass'' and circular velocity are
given by
\begin{equation}
\m178= {{4 \pi} \over{3}} \, 178 \, \bar\rho \, \r178^3,
\end{equation} and
\begin{equation}
\v178 = \left( {{G \m178} \over { \r178}} \right)^{1/2} .
\end{equation} These quantities prove to be useful for characterising
the global properties of realistic dark matter haloes.

\begin{figure}
\centering
\centerline{\epsfxsize= 10 cm \epsfbox[0 48 400 730]{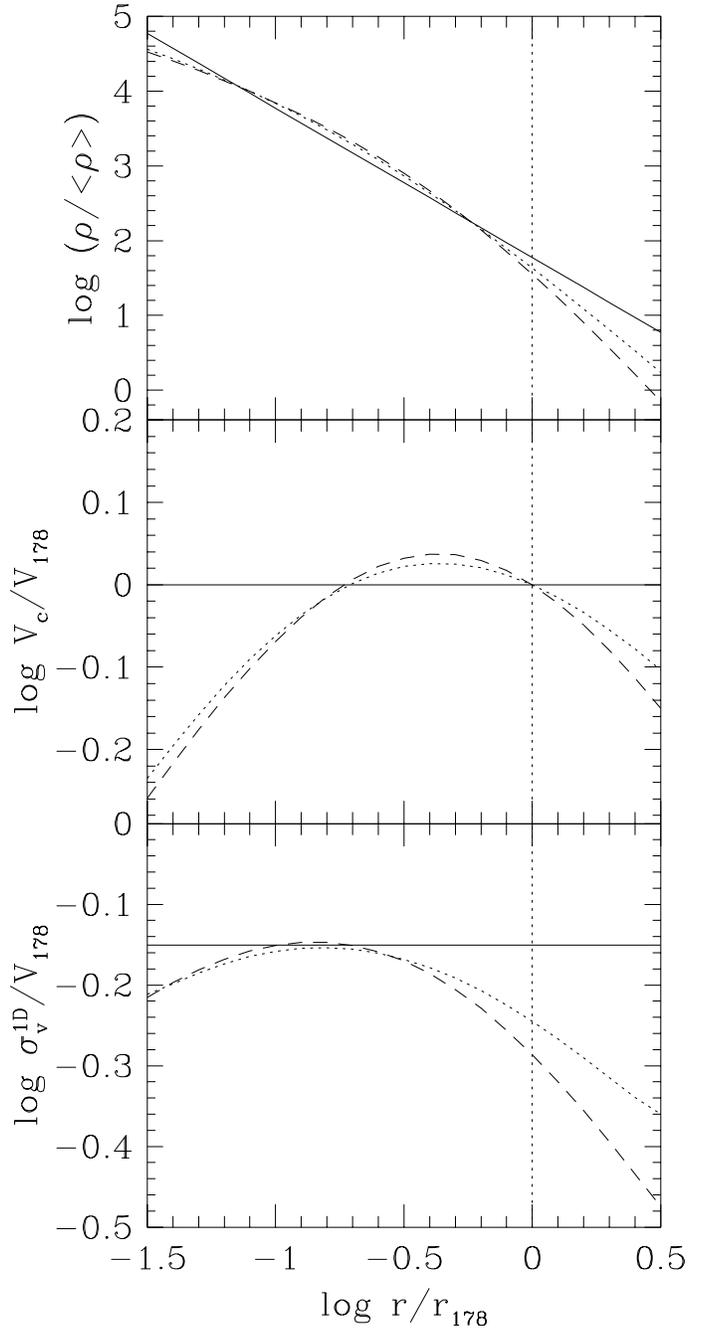}}
\caption{Comparison of the density, $\rho$, circular velocity,
$\protect{\vc}$, and velocity dispersion, $\protect{\s1d}$, profiles
for the singular isothermal sphere (solid) and the isotropic Hernquist
(dashed) and
\protect{\nfw} (dotted) models. In the \protect{\nfw} model the
parameter $\protect{\an} = 0.2$, as advocated to model the density
profile of a CDM galaxy cluster by Navarro \etal (1995a), and
$\protect{\ah} = 0.43$ in the Hernquist model so that both have
circular velocity profiles that peak at $r \approx 0.43
\protect{\r178}$. The ``virial radius'' $\r178$ is marked by the vertical
dotted line. }
\label{fig:models}
\end{figure}

Below we describe three equlibrium models for the internal structure
and kinematics of DM haloes, which we will later compare with the
properties of the haloes formed in a set of large cosmological N-body
simulations. These models are the singular isothermal sphere, the
Hernquist \shortcite{hq} and the Navarro, Frenk \& White
\shortcite{nfwa} (hereafter NFW) models. The latter two models differ
from the isothermal sphere in having shallower density gradients in
the inner parts and steeper gradients in the outer parts. All three
models are intended to represent only the regions interior to the
virial radius $\r178$. The Hernquist model has finite total mass,
while isothermal sphere and NFW models both have infinite total
mass. The Hernquist and NFW models can both be considered, in some
sense, as only minor modifications of the isothermal sphere, as
interior to the virial radius, the density $\rho$ is approximately
proportional to $r^{-2}$ over the region containing most of the
mass. The density, circular velocity and velocity dispersion profiles
of these models are compared in Fig.~\ref{fig:models} for particular
values of the scale parameters.


\subsection{The Singular Isothermal Sphere}\label{sec:sis}

A standard model for dark matter haloes, motivated by flat rotation
curves, is the singular isothermal sphere, for which the density
profile is
\begin{equation}
\rho(r) \propto \frac{1}{r^2} .
\end{equation} 
Defining $\rd \equiv r/\r178$ as the radial distance in units of the
virial radius, we can rewrite this as
\begin{equation}
\frac{\rho(\rd)}{\bar\rho} = \frac{178}{3} \, \frac{1}{\rd^2} .
\end{equation} 
The corresponding expression for the mass within radius $r$, scaled to
the value at the virial radius, is
\begin{equation}
\frac{M(\rd)}{\m178} = s
\end{equation} 
The circular velocity is constant with radius,
\begin{equation}
\frac{\vc(\rd)}{\v178} = 1,
\end{equation} 
and, assuming dynamical equilibrium and an isotropic velocity distribution,
the 1-dimensional velocity dispersion is also constant,
\begin{equation}
\frac{\s1d(\rd)}{\v178} = \sqrt{2} ,
\label{eq:sigma_iso}
\end{equation} 
In their relatively low resolution CDM simulations, Frenk
\etal (1985,1988) and Quinn, Salmon \& Zurek (1986) found halo
circular velocities which were essentially constant with radius down
to the gravitational softening radius, consistent with this simple
model.  This model, truncated at the virial radius, been widely used
as a description of the dark matter haloes formed in hierarchical
clustering models, \eg in the modelling of gravitational lensing
statistics by Narayan \& White (1988), and in the galaxy formation
models of Kauffmann, White \& Guiderdoni (1993) and Cole
\etal (1994).

\subsection{The Hernquist Model}

Hernquist \shortcite{hq} presented a completely analytical model which
closely approximates the de Vaucouleurs $R^{1/4}$ law that has long
been used to fit the surface brightness profiles of elliptical
galaxies. The density profile has the form
\begin{equation}
\rho(r) \propto \frac{1}{r(r+\bh)^3}
\end{equation}
Defining a dimensionless radius $\rd \equiv r/\r178$ as before, and a
dimensionless scale radius $\ah\equiv \bh/\r178$, the density, mass,
potential $\psi$ and circular velocity can be written in dimensionless form as
\begin{eqnarray}
\frac{\rho(\rd)}{\bar\rho} &=& 178\frac{2}{3} \, \frac{\ah(1+\ah)^2}{ \rd
(\rd+\ah)^3}
\\
\frac{M(\rd)}{\m178} &=& \frac{(1+\ah)^2 \rd^2}{(\rd+\ah)^2}
\\
\frac{\psi(\rd)}{\v178^2} &=& - \frac{(1+\ah)^2}{(\rd+\ah)} 
\\
\frac{\vc(\rd)}{\v178} &=& \frac{(1+\ah) } {(\rd+\ah)} \, \rd^{1/2},
\end{eqnarray} 
The parameter $\ah$ sets the scale at which the slope of the density
profiles changes from $r^{-1}$ at small $r$, to $r^{-4}$ at large
$r$. The circular velocity profile peaks at $r= \ah \r178$ (see
Fig.~\ref{fig:models}).  The main attraction of this model is that it
has an analytical distribution function which can be expressed in
terms of elementary functions.  It has also been used by Dubinski \&
Carlberg (1991) to fit the density profiles of galactic haloes formed
in their N-body simulations.

Assuming dynamical equlibrium, the radial velocity dispersion,
$\sigma_r(r)$, of the model can be obtained by integrating the Jeans
equation
\begin{equation}
\frac{1}{\rho} \frac{d}{dr} (\rho \sigma^2_{\rm r}(r) ) + 2 \beta
\frac{\sigma^2_{\rm r}(r)}{r} = - \frac{d\psi}{dr} ,
\label{eq:jeans}
\end{equation} 
where $\beta \equiv 1 -\sigma^2_\theta(r)/\sigma^2_{\rm
r}(r) $.  For the case of isotropic orbits, the radial and tangential
velocity dispersions are equal, $\sigma_\theta(r)=\sigma_{\rm r}(r)
$. For this case Hernquist \shortcite{hq} finds
\begin{eqnarray}
\frac{\sigma^2_{\rm r}(\rd)}{\v178^2} =  \frac{(1+\ah)^2}{(12\ah)} \biggl\{
\frac{12 \rd (\rd+\ah)^3}{\ah^4}\ln\left(\frac{\rd+\ah}{\rd}\right)
\nonumber - \\ \frac{\rd}{\rd+\ah}
\left[ 25 + 52 \left(\frac{\rd}{\ah}\right) + 42
\left(\frac{\rd}{\ah}\right)^2 + 12 \left(\frac{\rd}{\ah}\right)^3
\right] \biggr\}.
\end{eqnarray}


\subsection{The \nfw Model}

Navarro \etal (1995a) proposed an alternative analytical density
profile, which they found to be an excellent fit to the dark matter
haloes of galaxy clusters formed in their CDM simulations.  Navarro
\etal (1995b) found this profile to apply over a large range of masses
from those of dwarf galaxies to galaxy clusters. The density profile
has the form
\begin{equation}
\rho(r) \propto \frac{1}{r(r+\bn)^2}
\end{equation}
Defining $\an\equiv \bn/\r178$, the density, mass,
potential and circular velocity can be written as
\begin{eqnarray}
\frac{\rho(\rd)}{\bar\rho} &=& \frac{178}{3} \, \frac{f(\an)}{\rd \,
(\rd+\an)^2} 
\\
\frac{M(\rd)}{\m178} &=& f(\an) \,
\left[\ln\left({1+\frac{\rd}{\an}}\right)
- \frac{\rd}{(\rd+\an)} \right]
\\
\frac{\psi(\rd)}{\v178^2} &=& - f(\an) \, \frac{ \ln(1+\rd/\an) }{\rd} 
\\
\frac{\vc(\rd)}{\v178} &=& \frac{f(\an)^{1/2}}{\rd^{1/2}}
\left[ \ln\left({1+\frac{\rd}{\an}}\right)
- \frac{\rd}{(\rd+\an)} \right]^{1/2}
\end{eqnarray} 
where, for compactness, we have defined
\begin{equation}
f(\an) = \frac{1}{\left( \ln\left({1+1/\an}\right)
- 1/(1+\an) \right)} .
\end{equation}
Like the Hernquist profile, $\rho \propto r^{-1}$ as $r \rightarrow
0$, but the asymptote at large $r$ is now $r^{-3}$ rather than
$r^{-4}$, which leads to a logarithmically divergent total mass. The
parameter $\an$, again, defines the scale of the transition between
the inner and outer slopes, which occurs more gradually in this model
than in the Hernquist model (see Fig.~\ref{fig:models}). The slope of
the density profile is $d\ln\rho/d\ln r = -2$ at $r=\an\r178$. The
circular velocity peaks at $r=2.16\an\r178$. Also shown in
Fig.~\ref{fig:models} is the 1-dimensional velocity dispersion,
$\s1d$, obtained by numerically integrating equation~(\ref{eq:jeans}),
again assuming isotropic orbits and dynamical equilibrium.


\section{Simulations and Groups}\label{sec:sims}

\subsection{Simulations}

The simulations we have analysed are the three scale-free simulations
which were used to study the statistics of mergers in an $\Omega=1$
universe by Lacey \& Cole \shortcite{lc2}.  The simulations were
performed using the high resolution particle-particle-particle-mesh
($P^3M$) code of Efstathiou \etal \shortcite{edfw} with $128^3\approx
2 \times 10^6$ particles. The long-range force was computed on a
$256^3$ mesh, while the softening parameter for the short-range force
was chosen to be $\eta=0.2(L/256)$, where $L$ is the size of the
(periodic) computational box. The corresponding potential can be
approximated by a Plummer law with softening $\epsilon \approx
\eta/3 = L/3840$. The initial power spectra of the simulations were power 
laws, $P(k) \propto k^n$, with $n=-2$,$-1$ and~$0$. More details are
given in Lacey \& Cole \shortcite{lc2}.

In this paper we analyse almost exclusively the last output from each
of the three simulations. At this output the three simulations have
expanded by a factor $27.8$, $13.3$ and $6.3$ and the characteristic
mass is $\mstar=266$, $447$, and $46.8$ particles for $n=0$, $-1$
and~$-2$ respectively.  Here we define the characteristic mass,
$\mstar$, such that the r.m.s. linear density fluctuation in spheres
containing mass $\mstar$ is $1.69$ -- this being the linear theory
density corresponding to the collapse to a singularity of a uniform
spherically symmetric perturbation.  To test for the effects of
resolution we occasionally analyse a set of earlier outputs from these
simulations at which the values of $\mstar$ were a factor of 4
smaller.


\subsection{Group Finders}

We wish to study the intrinsic properties of non-linear self-bound
structures that form via gravitational instability in the N-body
simulations. For this reason we have identified these
structures using different group finding algorithms to ensure that the
properties we quantify are intrinsic to the structures and not
artifacts of any particular group finding algorithm.  Thus to identify
the groups in the N-body simulation we employed two different group
finding algorithms, the standard friends-of-friends (hereafter FOF)
algorithm of Davis \etal (1985) and the spherical overdensity
(hereafter SO) algorithm described in Lacey \& Cole
\shortcite{lc2}.

FOF groups are constructed by linking together all pairs of particles
whose separation is less than $b$ times the mean inter-particle
separation. This results in groups bounded by a surface of
approximately constant density, $\rho / \bar \rho \approx 3/(2 \pi
b^3)$. This elegantly simple algorithm has been used extensively in
previous analyses of N-body simulations, usually with $b=0.2$ (\eg
Frenk \etal 1988, Efstathiou \etal 1988).  It succeeds in picking out
most of the groups one can identify by eye, but occasionally can join
together two or more distinct density centres that are linked by a
tenuous bridge of particles. We will denote groups defined using this
algorithm by FOF($b$) .

The SO algorithm first ranks all the particles in the simulation by
their local density, computed from the distance to their 10th nearest
neighbour.  Then, starting with the densest, a sphere is grown around
this centre until the enclosed density drops below some threshold,
$\kappa\bar\rho$. The group centre is redefined to be the centre of
mass of the selected region and the process of growing the sphere and
selecting a new region repeated iteratively until the group centre and
group membership converge. Particles assigned to a group in this way
are removed from the density ranked list and not considered
further. Then in the same manner each of the remaining ungrouped
particles in the ranked list are used as group centres. Finally, small
groups inside or overlapping with larger groups are merged into the
larger object.  This algorithm selects spherical regions whose average
density is equal to $\kappa\bar\rho$. A similar method, with a mean
overdensity of 180, has been used by Warren \etal (1992). We denote
groups found in this way by SO($\kappa$).


\subsection{Halo Profiles}

In constructing halo profiles we have chosen to include all the
particles surrounding the group and not only those assigned to the
group by the group finding algorithm.  In each of the groups defined
above we locate the particle with the most negative gravitational
potential energy. The potential is calculated using only the group
particles, and assuming that the potential due to each particle is
$-1/(r^2+\epsilon^2)^{1/2}$, with $\epsilon=\eta/3$. We then use this
as the centre of the group and inflate a sphere around this centre
until the enclosed density drops below $178 \bar \rho$. The radius of
this sphere and the mass it encloses define $\r178$ and $\m178$ for
the group. We use $\r178$, $\m178$ and the corresponding $\v178$ to
scale the halo profiles. These choices allow a fair comparison of
groups of different masses and groups located using different
algorithms.  The question of which algorithm best divides the
simulation into discrete virialized groups then becomes largely a matter 
of which returns a group mass close to the virialized mass, which we will 
see is close to our adopted definition of the virial mass $\m178$.  
The relation between $\m178$ and the original group mass, $M_{\rm group}$ 
is studied in Section~\ref{sec:bulk}.

\begin{table}
\caption{ The distribution of SO(178) groups by mass in the three
simulations.  The last line gives the mass $\mstar$ in terms of the
number of particles.}
\begin{center}
\begin{tabular}{rrrrrrrrrr}
\hline
\multicolumn{1}{l} {Spectral Index } &
\multicolumn{1}{r} {$n=0$} &
\multicolumn{1}{r} {$n=-1$} &
\multicolumn{1}{r} {$n=-2$} \\ $\m178/M_*$ & & & \\
\hline 32--64\hphantom{.}  & -- & -- & 16 \\ 16--32\hphantom{.}  & --
& -- & 72 \\ 8--16\hphantom{.}  & 1 & 14 & 134 \\ 4--8\hphantom{2.}  &
29 & 42 & 310 \\ 2--4\hphantom{2.}  & 222 & 141 & 649 \\
1--2\hphantom{2.}  & 692 & 282 &1280 \\ 0.5--1\hphantom{2.}  &1563 &
592 &2486 \\ 0.25--0.5 &2832 & 1109&4786 \\ & & & \\ $\mstar$ &266
&447 &46.8 \\
\hline
\end{tabular}
\end{center}
\label{tab:masses}
\end{table}

For each group centre we estimate a comprehensive selection of the
properties of the surrounding material binned both in spherical shells
and cumulatively within spheres. These properties include: the
density, circular velocity, radial velocity, radial and tangential
velocity dispersions, rotation velocity, angular momentum, axial
ratios of the moment of inertia tensor and virial ratio. All internal
velocities are computed relative to the centre of mass velocity of the
sphere of radius $\r178$. The radial bins were chosen to be equally
spaced for the innermost 10~bins and then with logarithmically
increasing widths out to an outer radius of $10\r178$. The spacing of
the inner bins was chosen such that the innermost bin, which in
general is the least populated, typically contained 10~particles.
When computing profiles averaged over all haloes within a mass range,
the bin radii were scaled with $\r178$. We examine individual profiles
constructed in this way, and also mean profiles of the scaled
quantities ($\rho/\bar\rho$, $V_c/\v178$ etc.) averaged over groups
within mass bins. The mass bins were chosen to cover a factor 2 in
mass.  With the exception of the density profile, the mean profiles
evaluated in spherical shells are weighted means. We gave equal weight
to all shells in the individual profiles that contained 10 or more
particles and zero weight to those containing fewer than 10~particles.
This procedure was adopted to avoid low particle numbers producing a
large scatter in profiles such as the velocity dispersion as a
function of radius.  For each of the mean profiles we also accumulated
the group-to-group dispersion at each radius.  These dispersions can
be used to place errorbars on each of the profiles which reflect the
group-to-group variation.

\begin{figure*}
\centering
\centerline{\epsfxsize= 18cm \epsfbox[0 320 574 690]{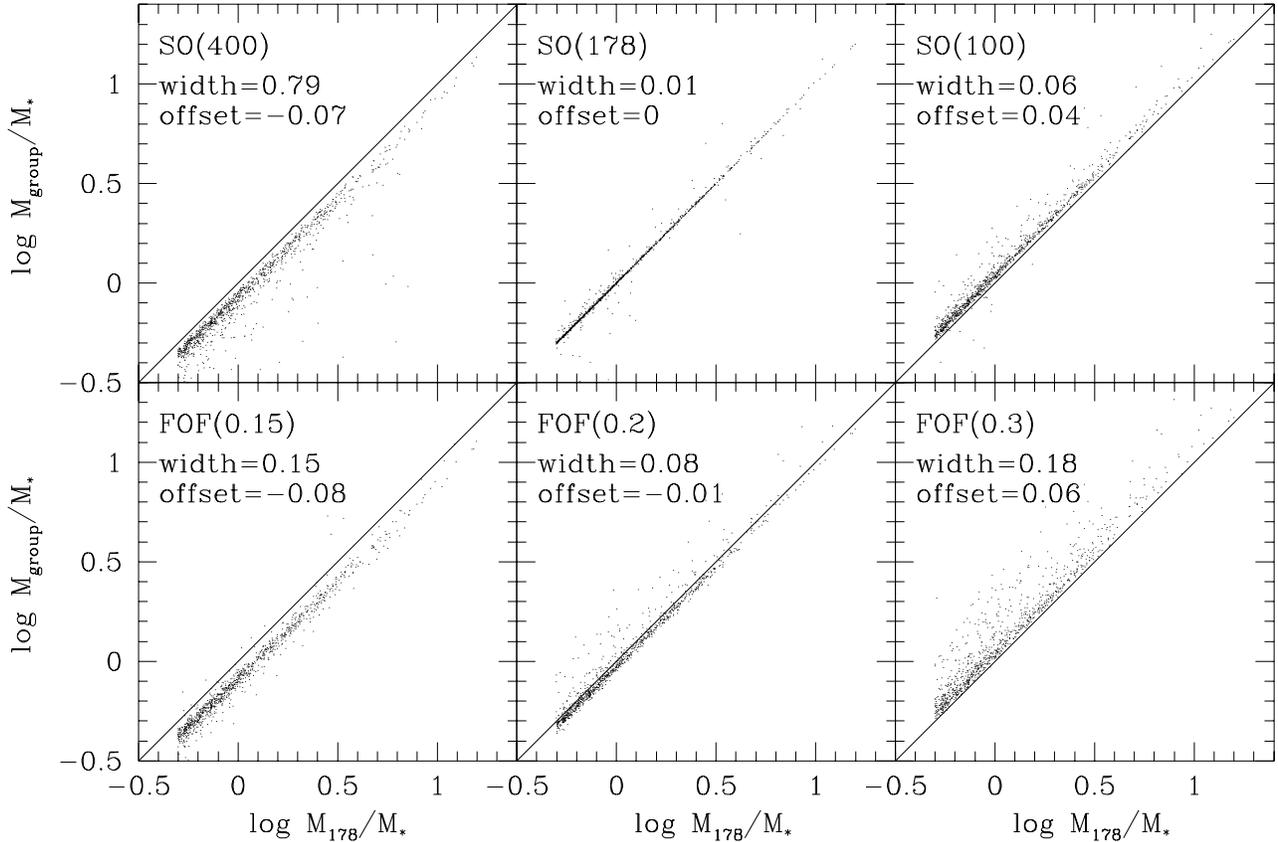}}
\caption{Scatter plots of $\mg$ versus $\m178$ for groups identified
in the $n=-1$ simulation using each of the group finders SO(400),
SO(178), SO(100), FOF(0.15), FOF(0.2) and FOF(0.3). 
The mass $\mstar$ equals 447 particles for this simulation.
The straight
indicates the locus $\mg=\m178$. This line is omitted from
the SO(178) plot where it would obscure most of the data points. The
offset given in each panel is the median of ${\rm
log}_{10}(\mg/\m178)$. Also specified is the width of the
distribution in ${\rm log}_{10}(\mg/\m178)$ measured between the
10th and 90th centiles.  }
\label{fig:scatm}
\end{figure*}

\begin{figure*}
\centering
\centerline{\epsfxsize =18 cm \epsfbox[0 320 574 690]{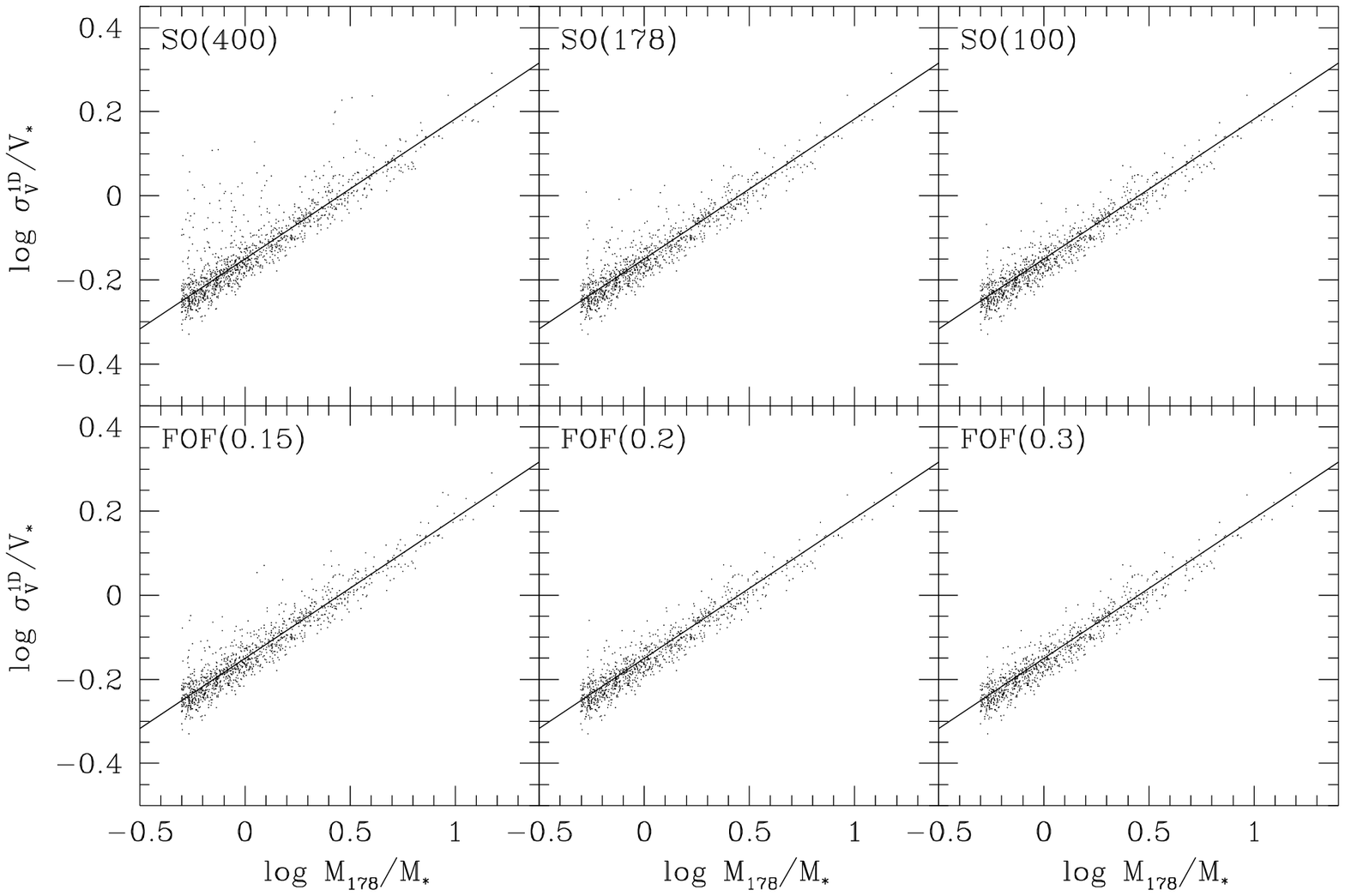}}
\caption{Scatter plots of mean 1-dimensional velocity dispersion,
$\s1d$, within the virial radius, $\r178$, versus the mass $\m178$ for
groups identified in the $n=-1$ simulation using each of the group
finders SO(400), SO(178), SO(100), FOF(0.15), FOF(0.2) and
FOF(0.3). The mass $\mstar$ equals 447 particles for this simulation.
$V_*$ is $\v178$ for a halo with $\m178=\mstar$. The
straight indicates the locus $\s1d = \v178/2^{1/2}$, which is the
prediction of the isothermal sphere model.}
\label{fig:scatv}
\end{figure*}

Table~\ref{tab:masses} gives the distribution of masses in each of the
three simulations for centres identified by SO(178).  The large
numbers of groups enable the mean profiles in each mass range to be
determined very precisely. This then enables the dependence of halo
profiles on mass and spectral index to be examined in some detail, but
care must be taken to assess how resolution, which in terms of
$\r178$ worsens with decreasing $\m178/\mstar$, affects these
profiles.  It is for this reason that we will also examine the
profiles of groups extracted from an earlier output of the simulations
when $\mstar$ was a factor 4 smaller.

\section{Bulk Properties}\label{sec:bulk}

  We first look at the basic global properties of individual groups
selected using the various group finders.  As we will see in
Section~\ref{sec:sphere}, when we study the dynamical structure of the
haloes, the radius $\r178$ corresponds well with the boundary at which
there is a transition from general infall to approximate dynamical
equilibrium. Thus the virial mass of the halo is well approximated by
$\m178$. It is therefore interesting to see how the group mass defined
by the various algorithms compares with $\m178$, as this acts as a
measure of how well the particular group finders succeed in
partitioning the simulation into discrete virialized systems.  We see
in Fig.~\ref{fig:scatm} that for each group definition the group mass,
$M_{\rm group}$, correlates strongly with $\m178$, but the offset of
the correlation and the scatter around it vary. The offset between $M_{\rm
group}$ and $\m178$ is easily understood.  Groups found using high
$\kappa$ or small $b$ are typically denser and less massive than those
found using small $\kappa$ or large $b$, and so they are the central
regions or sub-clumps of the less dense groups.  Note that for FOF the
offset is smallest for FOF(0.2).  The scatter in each relation is of
more interest.  As is to be expected from the similarity of their
definitions, the scatter between $\mg$ and $\m178$ is smallest for
SO(178) groups.  In this case, the essential difference in the
definitions of $\mg$ and $\m178$ is the choice of group centre. For
SO(178) the group is centred by its centre of mass, while $\m178$ is
evaluated around the centre defined by the particle in the SO(178)
group with the most negative gravitational potential energy, where the
potential energy is evaluated using only the group particles.  This
difference is generally not significant as can be seen by the fact
that many of the points fall exactly on the line $\mg=\m178$.  For SO
groups identified with other values of the density contrast, $\kappa$,
the scatter grows. For the dense SO(400) groups there is a tail of
objects with $\m178 \gg \mg$, for which the SO(400) group is simply
the core of a larger virialized halo.  For FOF(0.2) or (0.3) the
scatter about the mean correlation is asymmetric with a tail of haloes
with $\mg > \m178$.  As the linking length $b$ is increased this tail
becomes more pronounced, extending to $\mg > 2\m178$.  These outliers
arise when two or more distinct density centres become linked by
tenuous bridges of particles.

Fig.~\ref{fig:scatv} shows the correlation of the 1-dimensional
velocity dispersion averaged within $\r178$ with the mass $\m178$.
The correlation agrees well with the expectation of the isothermal
sphere model (Section~\ref{sec:sis}), $\s1d = \v178/\sqrt{2}\propto
\m178^{1/3}$, with an r.m.s. scatter about this mean relation of only
10--15\%. For the groups selected at high density contrast,
particularly SO(400), a tail of groups appears with $\s1d >
\v178/\sqrt{2}$. In these cases, it seems likely that the sphere of
radius $\r178$ is actually only part of a significantly larger
virialized structure.

Groups selected at low density contrast, particularly those located by
FOF, tend to be composities of distinct virialized structures, while,
on the other hand, groups identified at higher density contrast are
often only small clumps within larger structures.  Overall the
FOF(0.2) and SO(178) group masses and $\m178$ correlate well. Thus
both group finding algorithms are good candidates for selecting
virialized structures. The disadvantage of the SO(178) algorithm is
that it forces spherical boundaries on the groups when we shall see
that groups are typically triaxial in shape.  The FOF(0.2) algorithm
does not force any particular geometry on groups, but because it is
only sensitive to the local density it is prone to occasional merging
of separate virialized systems that are linked by tenuous bridges.


\begin{figure}
\centering
\centerline{\epsfxsize = 7.0 cm \epsfbox[0 50 300
750]{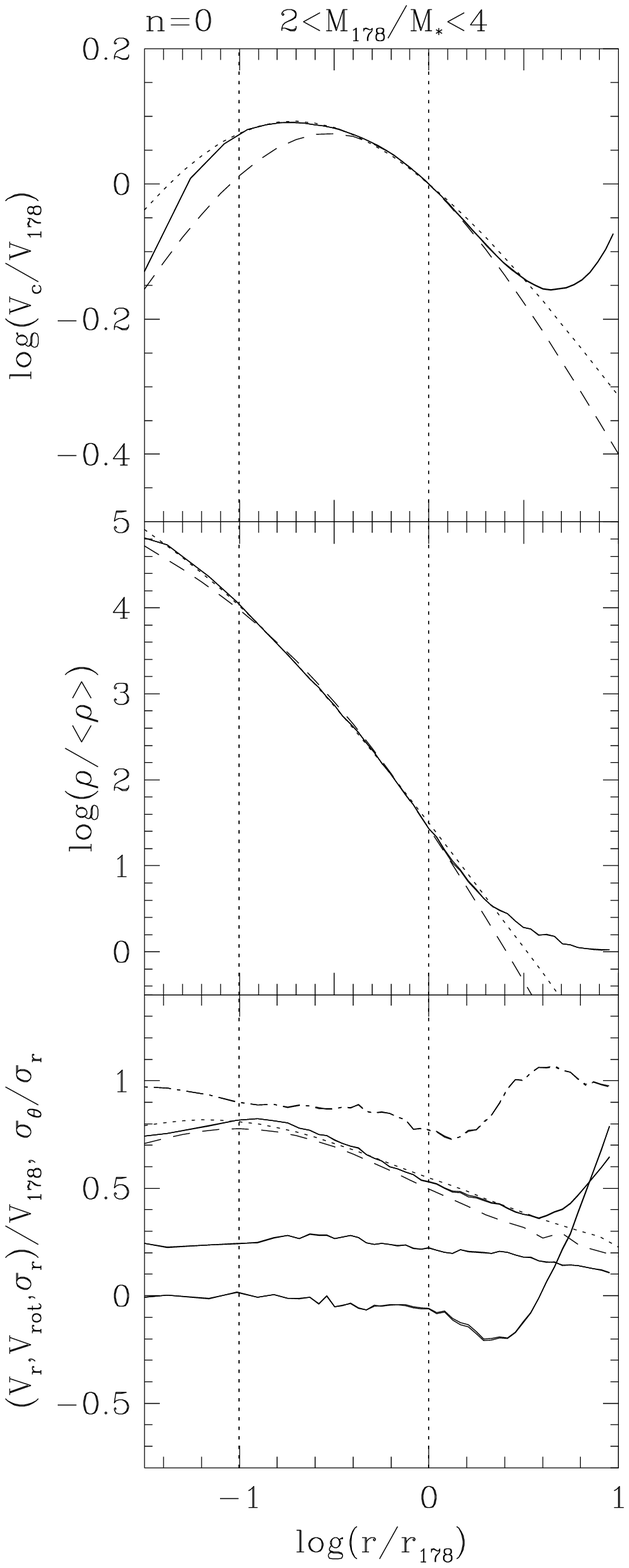}}
\caption{ The upper and middle panels show mean circular velocity and
mean density profiles of all groups with $2<\m178/M_*<4$ 
($532<N<1064$) 
in the
$n=0$ simulation located by both SO(178) and FOF(0.2).  In this case
the lists of group centres defined by SO(178) and FOF(0.2) were
identical. The smooth curves in these two panels are the Hernquist
(dashed) and \nfw (dotted) models which have been fitted to the
density profiles over the range between the vertical dotted lines
which mark the force-softening scale $\eta$ and virial radius $\r178$.
The lower panel shows a variety of measures of the dynamical state of
the haloes. The lowest solid curve is the mean radial velocity,
$V_{\rm r},$ as a function of radius. The next solid curve is the mean
rotation speed, $V_{\rm rot}$. The upper dot-dashed curve is
$\sigma_\theta/\sigma_{\rm r}$, which measures the anisotropy of the
velocity dispersion. The remaining solid curve traces the mean radial
velocity dispersion as a function of radius.  The smooth dashed and
dotted curves are respectively the Hernquist and
\nfw models for the velocity dispersion assuming isotropy,
$\sigma_\theta/\sigma_{\rm r} =1 $.  }
\label{fig:profn0a}
\end{figure}

\begin{figure}
\centering
\centerline{\epsfxsize = 7.0 cm \epsfbox[0 50 300
750]{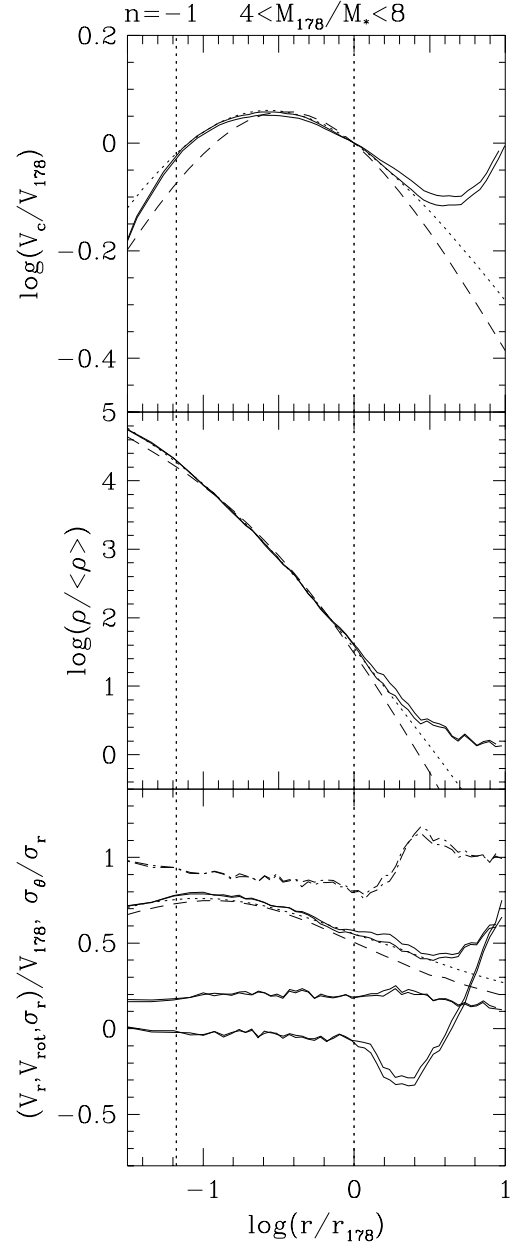}}
\caption{ As Fig.~\ref{fig:profn0a}, but for groups from the $n=-1$
simulation with masses in the range $4<\m178/\mstar<8$ 
($1788<N<3576$). 
 Here the lists
of group centres produced by the SO(178) and FOF(0.2) algorithms were
slightly different and thus two slightly different curves are visible
for each profile.  }
\label{fig:profnm1a}
\end{figure}

\begin{figure}
\centering
\centerline{\epsfxsize = 7.0 cm \epsfbox[0 50 300
750]{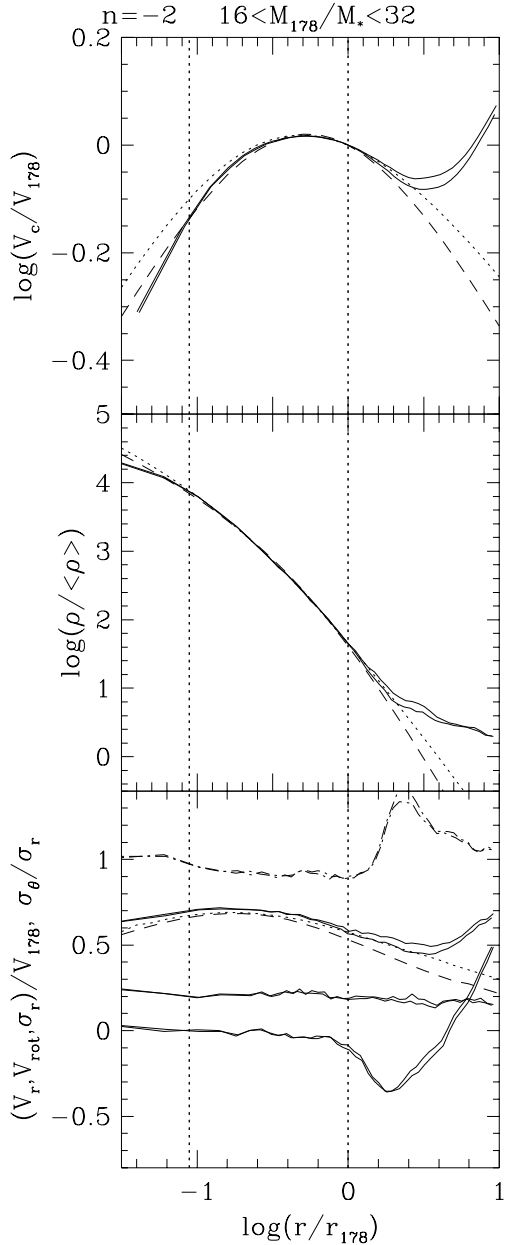}}
\caption{ As Fig.~\ref{fig:profnm1a}, but for groups from the $n=-2$
simulation with masses in the range $16<\m178/\mstar<32$
($749<N<1498$).  }
\label{fig:profnm2a}
\end{figure}

\begin{figure}
\centering
\centerline{\epsfxsize=9 cm \epsfbox[40 60 440 750]{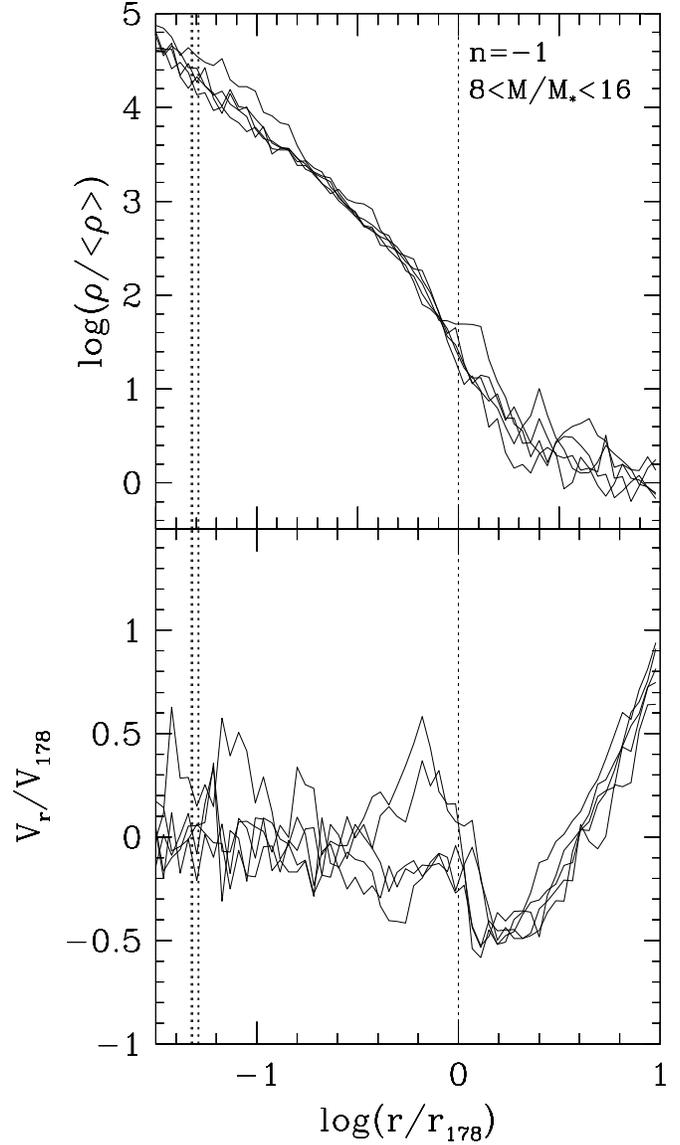}}
\caption{The density and radial velocity profiles of a selection of 5
groups taken from the $n=-1$ simulation with masses in the range
$8<\m178/\mstar<16$ 
($3576<N<7152$).  }
\label{fig:iprof5}
\end{figure}

\begin{figure}
\centering
\centerline{\epsfxsize=7.5 cm \epsfbox[95 430 415
730]{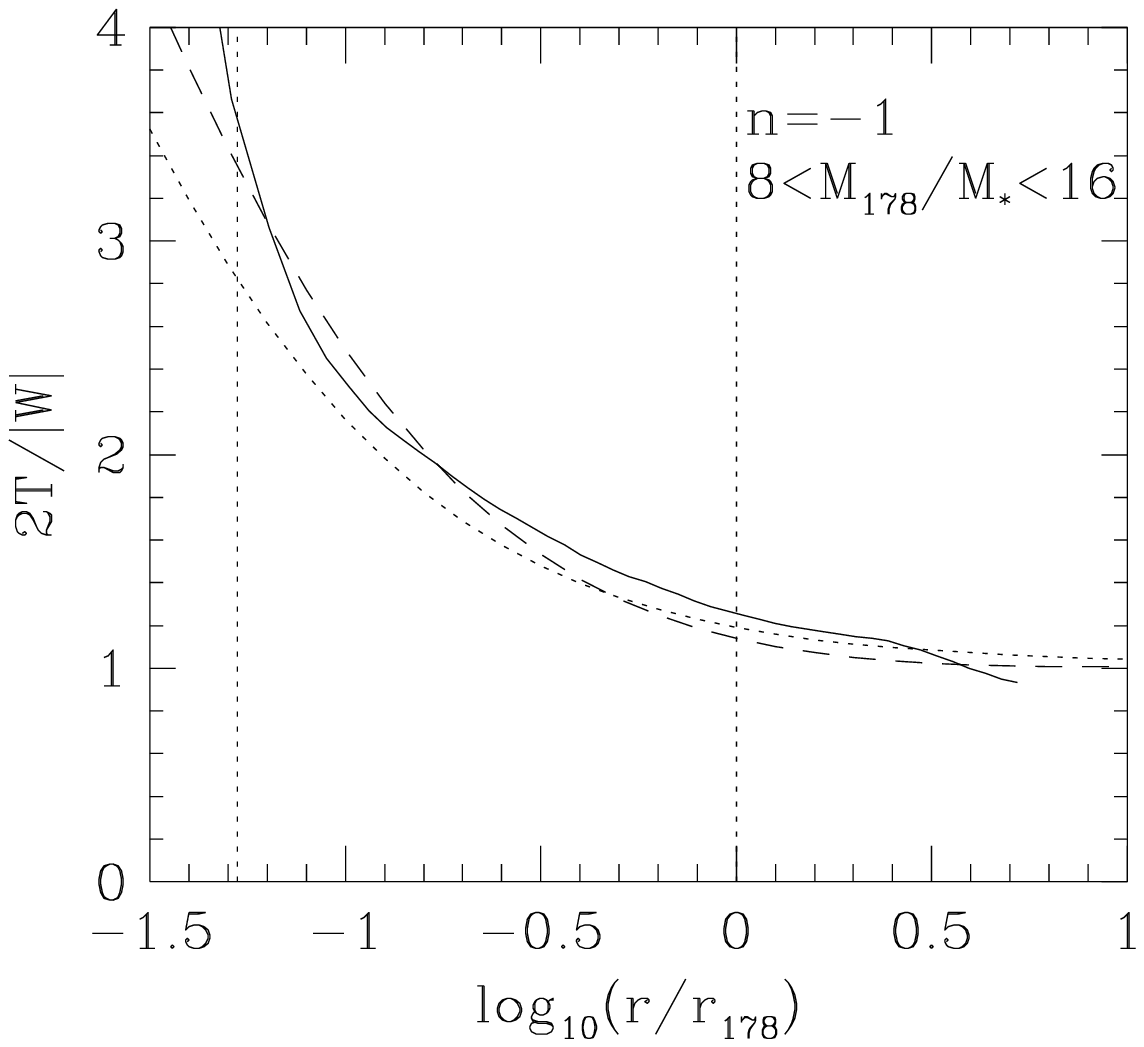}}
\caption{The solid curves shows the mean virial ratio, $2T/\vert W
\vert $, evaluated within spheres, as a function of radius, for haloes
with masses in the range $8<\m178/\mstar<16$ 
($3576<N<7152$) 
from the $n=-1$
simulation. The dashed and dotted curves show respectively the
corresponding profiles for the Hernquist and \nfw models that were
fitted to the mean density profiles of these groups.  }
\label{fig:prof_2tw}
\end{figure}

\section{Spherically Averaged Profiles}\label{sec:sphere}

We now examine the spherically averaged radial profiles of haloes and
compare them to the analytical models described in
Section~\ref{sec:models}.

\subsection{Massive Haloes}

Figs.~\ref{fig:profn0a}, \ref{fig:profnm1a} and~\ref{fig:profnm2a}
show various mean profiles (averaged over mass bins) around the
centres of the massive SO(178) and FOF(0.2) groups from each of the
three simulations.  Here, and in all subsequent plots of radial
profiles, two dotted vertical lines mark the force-softening scale
$\eta$ and virial radius $\r178$.  Note that the effective softening
for the $P^3M$ code is often taken be $\epsilon = \eta/3$. The upper
two panels show the mean circular velocity, $V_{\rm c}\equiv (
GM(<r)/r)^{1/2}$, and density profiles. The density profiles flatten
and the circular velocities turn up at large radii, $r \gg \r178$,
where material not physically bound or associated with the group
becomes included in the profile. Thus the density tends to the mean
density at large $r$ and the circular velocity becomes proportional to
radius. Also shown on these figures are the corresponding Hernquist
and \nfw profiles which have been fitted to the SO(178) mean density
profile over the range $\eta< r<\r178$.  The fits were performed by
converting the group-to-group dispersion of the density profiles at
each radius into an effective error on the mean profile. The scale
parameters $\ah$ and $\an$ of the Hernquist and \nfw models were then
varied to achieve minimum-$\chi^2$ fits. Visually, both models are
excellent fits to the halo density profiles between the virial radius,
$\r178$, and the force-softening scale $\eta$, for each value of
$n$. The \nfw model in particular continues to be a good match to the
halo profiles at larger radii, where we will see that the halo is not
yet in equilibrium, and at smaller radii where resolution is likely to
have affected the halo structure.  A more telling comparison is that
between the model and N-body circular velocity profiles. Note that by
definition these are all constrained to pass through the same point at
$\r178$.  We see that the Hernquist model changes slope too abruptly
to match that of the haloes. The \nfw profile is a significantly
better match.

The lower panels of Figs.~\ref{fig:profn0a}, \ref{fig:profnm1a}
and~\ref{fig:profnm2a} show various mean kinematic properties of the
haloes. These are the rotation speed $V_{\rm rot}$ in spherical shells
around the axis defined by the angular momentum vector of the shell,
the mean radial velocity $V_{\rm r}$, radial velocity dispersion
$\sigma_{\rm r} =\langle (V_{\rm r}-\langle V_{\rm
r}\rangle)^2\rangle^{1/2}$, and velocity dispersion anisotropy
$\sigma_\theta(r)/\sigma_{\rm r}(r)$, where $\sigma^2_\theta = \langle
V^2_{\theta}\rangle$ is the tangential velocity dispersion. In each
case the velocities are defined with respect to the centre of mass
motion of all the material interior to radius $\r178$.  To a first
approximation the rotation velocity is small and the velocity
dispersion isotropic and almost independent of radius, all in
agreement with the simple isotropic isothermal sphere model.  A more
detailed inspection shows that $V_{\rm rot}/\sigma_{\rm r} \approx
0.3$, which although appreciable, implies that the rotation is not a
significant contributor to the dynamical support of the halo.
Interior to $\r178$, $\sigma_\theta/\sigma_{\rm r}$ is slightly less
than unity, indicating a tendency for radial orbits, particularly at
$r\approx
\r178$.  Outside the virial radius $\sigma_\theta/\sigma_{\rm r}$
increases and slightly overshoots unity before settling to the
isotropic value, $\sigma_\theta/\sigma_{\rm r}=1$, at large radii.
Overlaying the velocity dispersion profiles are the model predictions
of the fitted Hernquist and \nfw models, computed assuming spherical
symmetry and $\sigma_\theta/\sigma_r=1$ ($\beta=0$), from the Jeans
equation~(\ref{eq:jeans}). We see that, within the virial radius, both
match well the slow curvature in the corresponding halo $\sigma_{\rm
r}$ profile.

The lowest curve in the lower panels of
Figs.~\ref{fig:profn0a},\ref{fig:profnm1a} and~\ref{fig:profnm2a} is
the mean radial velocity, $V_{\rm r}$ of material in shells.  These
curves are the main evidence we have that the radius $\r178$ is an
accurate characterisation of the virial radius of the haloes.  In each
case $V_{\rm r}$ is approximately zero interior to $\r178$, consistent
with a dynamically relaxed system. Between $\r178$ and $4\r178$
$V_{\rm r}<0$ indicating material in the process of falling onto the
central halo. At larger radii, $V_{\rm r}$ begins to increase passing
through $V_{\rm r}=0$ at the turn-around radius and matching onto the
general Hubble flow at large $r$.  The location of the turn-around
radius is in good agreement with the spherical collapse model which
predicts a mean density interior to the turn-around radius of $
(3\pi/4)^2\approx 5.55$. Similar results for the radial velocity were
found by Crone \etal (1994).

The typical group-to-group variation of the radial velocity pattern
and the corresponding density profiles is illustrated in
Fig.~\ref{fig:iprof5}, which shows the profiles for 5~groups taken
from the $n=-1$ simulation with masses in the range
$8<M/M_*<16$. Although the profiles can be significantly perturbed by
substructure in the individual haloes, the transition from infall to
quasi-static equilibrium at $r \approx \r178$ is quite clear in the
majority of cases. But the radial velocity profiles are too noisy to
make them useful for determining the boundary of the virialized region
for individual haloes.

Another measure of dynamical equilibrium within haloes is provided by
the virial ratio $2T/|W|$. The expectation for an isolated system in
dynamical equilibrium is for the total binding energy to equal twice
the total kinetic energy, $2T=|W|$.  As an example,
Fig.~\ref{fig:prof_2tw} plots the mean value of the virial ratio
$2T/|W|$ evaluated within spheres for groups from the $n=-1$
simulation with masses in the range $8<\m178/\mstar<16$. Here $T$ is
the kinetic energy of the material within a sphere of radius $r$ and
$W$ the self-gravitational binding energy of the same material, \ie
neglecting the the effect on the gravitational potential of all
material at larger radii. The ratio $2T/|W|$ should approach unity at
the boundary of the virialized region if surface terms in the virial
theorem vanish. We find $2T/|W|$ approaches unity only slowly at large
radii and is significantly greater than unity at $r=\r178$. However
this same profile is tracked quite well by the same quantity evaluated
from the fitted analytical Hernquist and \nfw models, which when
integrated over all the mass accurately obey the virial theorem.
Thus, although the haloes are believed to be in reasonably good
dynamical equilibrium within $r<\r178$, the ratio $2T/|W|$ is not
useful for defining the boundary of the virialized region.


\subsection{Mass dependent Trends}

\begin{figure*}
\centering
\centerline{\epsfbox{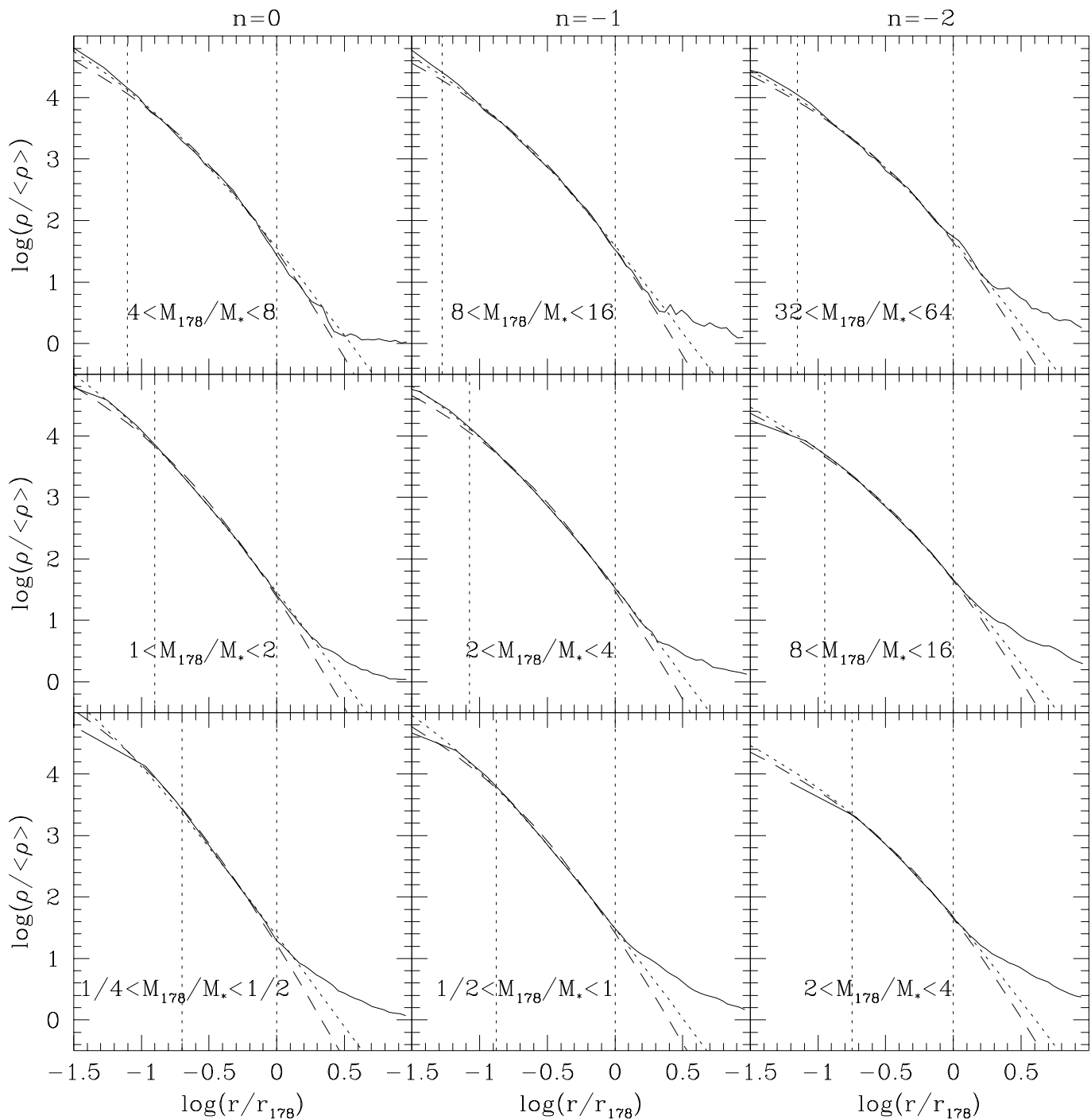}}
\caption{ The mean density profiles of SO(178) groups of a variety of
masses from each of the three self-similar simulations. The two dotted
vertical lines mark the force-softening scale, $\eta$, and virial
radius, $\r178$. The smooth curves are \nfw (dotted) and Hernquist
(dashed) model fits to the profiles in the range $\eta< r < \r178$.
Note that $\mstar$ equals $266$, $447$ and~$46.8$ in the $n=0$, $-1$ 
and~$-2$ simulations respectively.
}\label{fig:profrhos}
\end{figure*}

\begin{figure*}
\centering
\centerline{\epsfbox{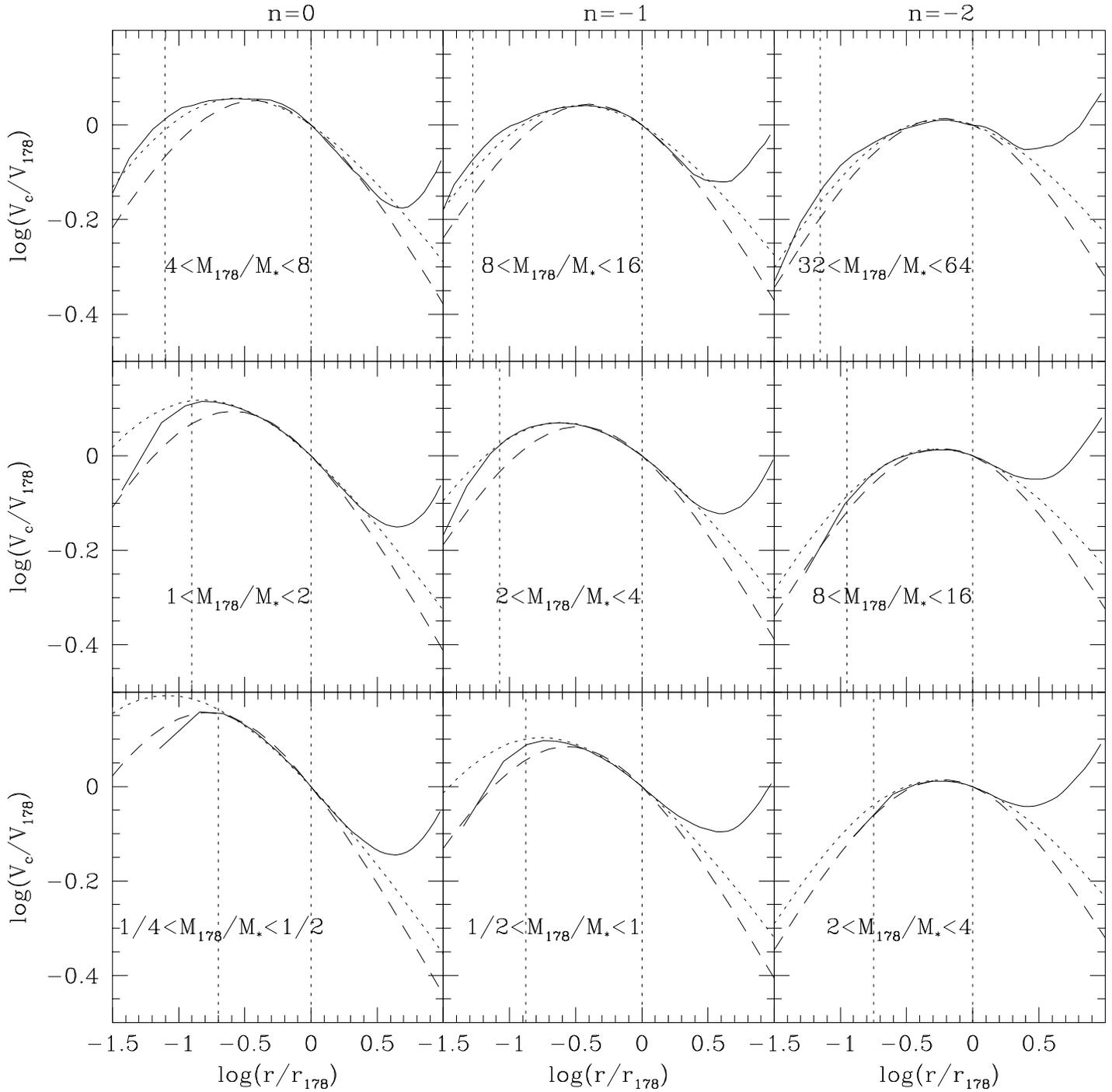}}
\caption{ This figure shows the circular velocity profiles and model
fits corresponding to Figure~\ref{fig:profrhos}.  }
\label{fig:profvcs}
\end{figure*}

In the previous section we saw that the \nfw model provides an
excellent description of the spherically averaged structure of the
more massive haloes formed in each of the three simulations.  We now
turn to how the halo structure depends on mass.  The density and
circular velocity profiles for a range of masses in each of the three
simulations are shown in Figs.~\ref{fig:profrhos}
and~\ref{fig:profvcs}.  Also shown are Hernquist and \nfw models which
have been fitted to the density profiles over the range
$\eta<r<\r178$.  Over this range, the \nfw model is an excellent
description of the halo density profiles for almost the full range of
masses investigated in each of the simulations.  Very close to the
softening radius $\eta$ the density profiles of the most massive halos
are slightly steeper than those of the \nfw model. For lower masses
the fitted \nfw models also match the inner portion of the halo
density profiles at $r<\eta$, in all cases except the low mass groups
of the $n=-2$ simulation.  Although not apparent to the eye, the
Hernquist models are formally worse fits to the halo density profiles
than the \nfw models. They tend to be marginally too steep in the
outer parts of the haloes and too shallow in the inner
regions. Nevertheless they are a considerable improvement over a
simple isothermal model, which has $\rho \propto r^{-2}$ at all radii.

\begin{figure*}
\centering
\centerline{\epsfbox{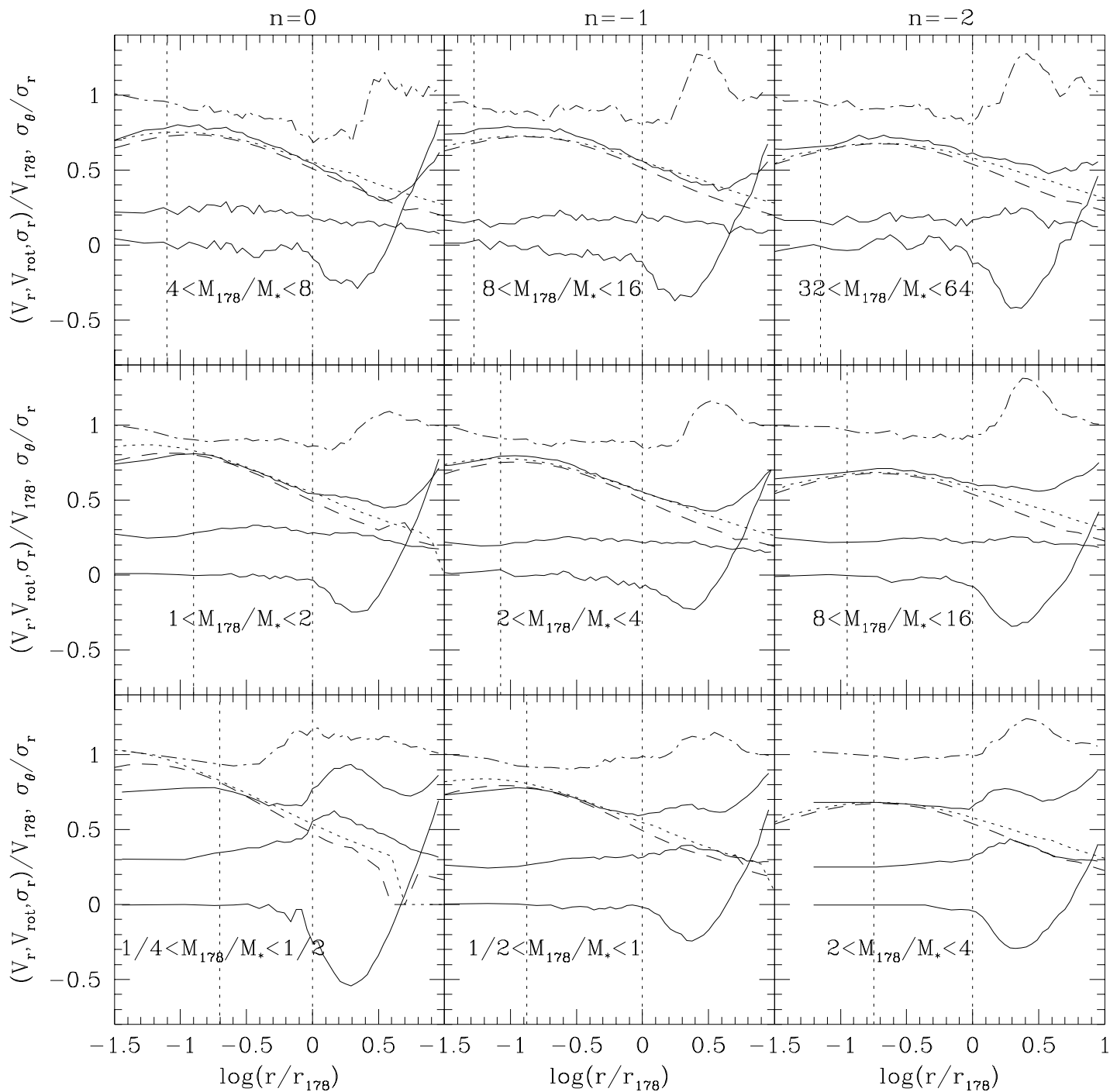}}
\caption{ This figure shows the mean profiles of kinematical properties
corresponding to Figure~\ref{fig:profrhos}. The lowest solid curve in
each panel is the measured radial velocity $V_r$, followed by rotation
speed $V_{rot}$ and radial velocity dispersion $\sigma_r$, and the
upper dot-dash curve is the velocity dispersion anisotropy. The smooth
dashed and dotted curves are the predictions of the isotropic
Hernquist and NFW models respectively, based on the fits to the
density profiles shown in Figure~\ref{fig:profrhos}. }
\label{fig:profvrs}
\end{figure*}

\begin{figure}
\centering
\centerline{\epsfxsize = 7.0 cm \epsfbox[0 50 300
750]{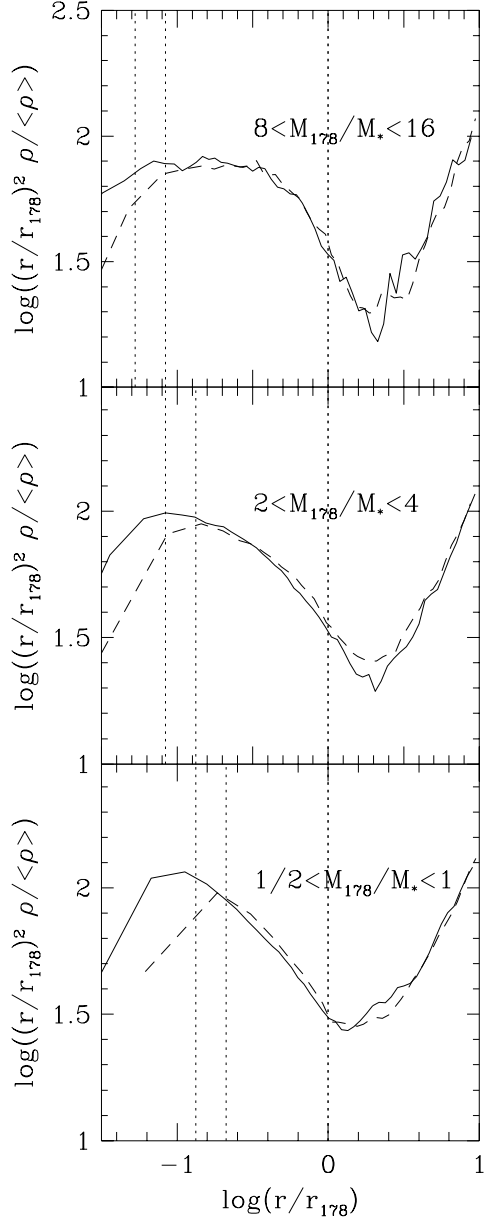}}
\caption{A comparison of the mean density profiles of SO(178) groups 
of a variety of masses in the $n=-1$ simulation at two different output 
times. The solid curves are the mean density profiles from the final output,
and are the same as those plotted in the central column of 
Fig~\ref{fig:profrhos} except here we plot 
$\log ( (r/\r178)^2 \rho/\langle \rho \rangle)$ so as to expand the
vertical scale. The dashed curves are the corresponding profiles from
an earlier output in which the groups are less well resolved. At this
earlier time $\mstar=112$ particles, a factor four less than at the 
final time. The inner two vertical dotted lines mark the force softening 
scale $\eta$ in the two cases.
}
\label{fig:soft}
\end{figure}

The differences between the \nfw and Hernquist models are more
pronounced in plots of the circular velocity profiles,
Fig.~\ref{fig:profvcs}.  Here we again see that the \nfw models match
well the profiles of the haloes over the entire region of the fit. In
contrast the Hernquist models peak too sharply and underestimate the
circular velocity in the central regions of the haloes.

\begin{figure}
\centering
\centerline{\epsfxsize = 9 cm \epsfbox{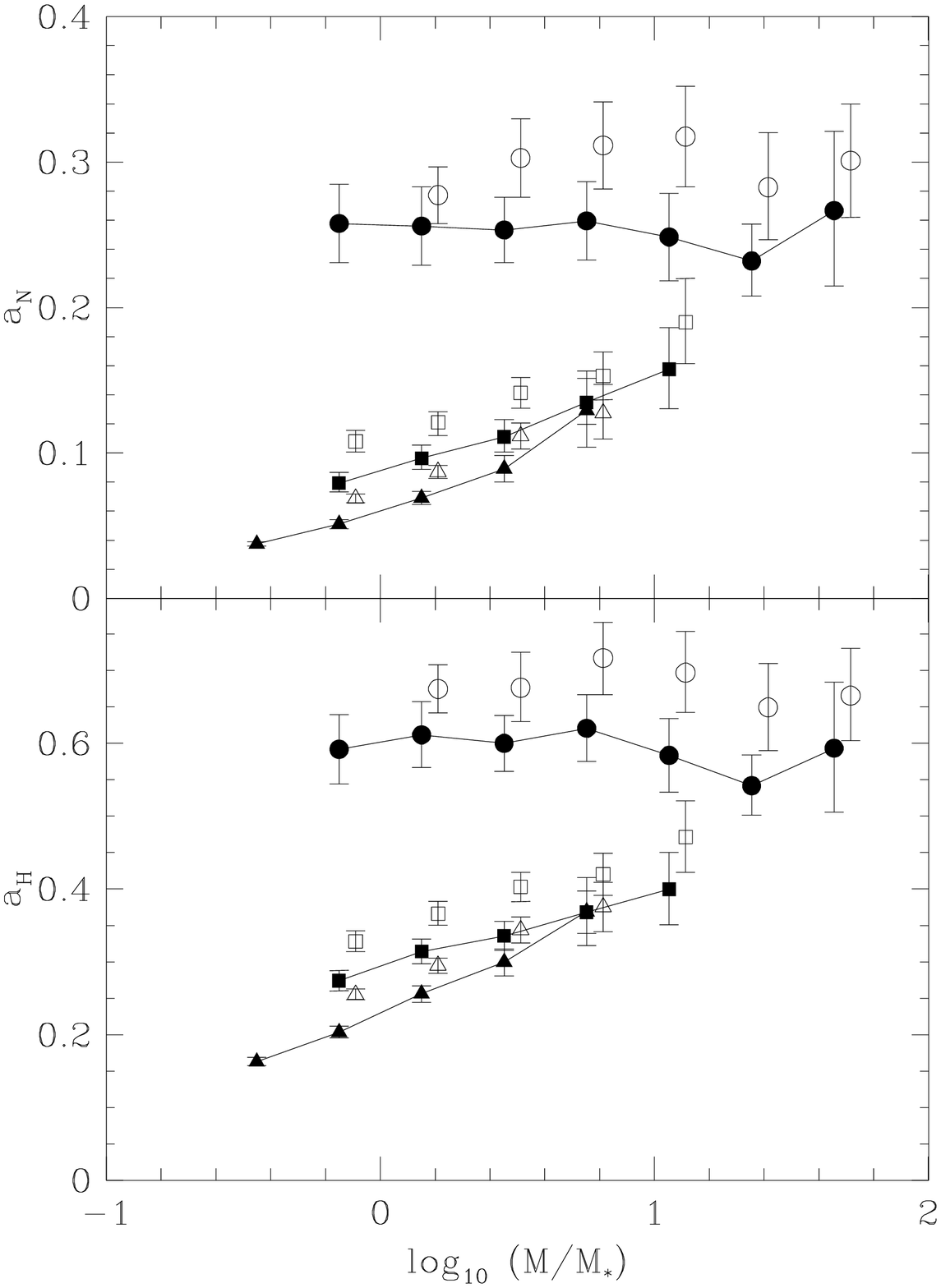}}
\caption{ The variation of the scale parameters $\protect{\an}$ and
$\protect{\ah}$ of the \nfw and Hernquist model fits to the halo
density profiles. The solid symbols are the parameter values from the
fits to the mean density profiles of the groups from the final output
of each simulation. The circles, squares and triangles are for $n=-2$,
$-1$ and $-2$ respectively. $\mstar$ for these outputs corresponds to
$266$, $477$ and $46.8$ particles. The open symbols, plotted displaced
slightly to the right, are the corresponding parameter values obtained
from earlier simulation outputs when the characteristics masses,
$\mstar$, were a factor~4 smaller. The error bars on each point
indicate the $\pm5\sigma$ range, where the formal $1\sigma$ errors
were obtained from the $\Delta\chi^2=1$ points of the $\chi^2$-fits to
the mean density profiles. }
\label{fig:afit}
\end{figure}

We now turn to the question of systematic variations of the halo
density profiles with mass, $\m178/\mstar$. Here we have to be very
careful to distinguish between real physical effects and the
consequences of the limited resolution of the N-body simulations. If
the finite number of particles or the force resolution were
artificially supressing the density  in the centre of the simulated haloes
then we would expect the radius of the affected region
to be directly proportional to $\eta$.  
In contrast we see in Fig.~\ref{fig:profvcs} that the peak in
the circular velocity profiles moves to smaller $r/\r178$ for lower
mass haloes while at the same time $\eta/\r178$, marked by the inner
vertical line, increases. This is the first piece of evidence which
gives us confidence that the form of the density and circular velocity
profiles is not predominately due to limited resolution. A more
stringent test comes from analysing the haloes with the same values of
$\m178/\mstar$, but extracted from an earlier output of each
simulation when $\mstar$ was smaller by a factor~4, and therefore
$\eta/\r178$ larger by a factor~1.59. An example of this comparison
is shown in Fig~\ref{fig:soft} which compares the mean density profiles 
of groups in the $n=-1$ simulation at the two output times.  
We note that the two density profiles agree quite accurately for radii 
greater than $\eta$, the force softening scale of the less well resolved
output. 
Fig~\ref{fig:afit} shows the
dependence of the scale parameters $\an$ and $\ah$ on $\m178/\mstar$
for each of the three simulations and for the two output times.  The
fits were made to the binned mean density profiles over the range
$\eta<r<\r178$ using estimates of the error on the mean profile
obtained from the group-to-group variance about the mean divided by
the number of groups. Standard $\chi^2$-fits were made, treating each
data point as independent and formal $1\sigma$ errors on the fit
parameters estimated from the points adjacent to the best fit where
$\Delta\chi^2=1$. The error bars on Fig~\ref{fig:afit} show the
$\pm5\sigma$ bounds.  If numerical resolution were the sole reason for
the existence of the detectable break in the halo density profiles,
then we would expect the scale lengths $\an$ or $\ah$ to scale as
$M^{-1/3}$ at a given output time, and we would expect the values from
the earlier output to be 60\% larger than those from the final
output. In fact they are typically less than 20\% larger than their
better resolved counterparts from the final output of each
simulation. Thus, although the measured scale lengths must have some
residual dependence on the numerical resolution, the detected trends
with mass and spectral index are real.  The scale lengths, $\an$ or
$\ah$, increase with increasing mass $\m178/\mstar$, and the steepness
of this relation depends on the spectral index $n$.

We also examined the mean profiles of velocity dispersion, anisotropy,
rotation velocity and radial velocity for haloes in all mass
ranges. The results, shown in Figure~\ref{fig:profvrs}, are very
similar for the different masses. In particular, the radial velocity
always starts to go significantly negative (indicating the edge of the
virialized region) at $r\approx\r178$.

\begin{figure}
\centering
\centerline{\epsfxsize=9 cm \epsfbox[90 414 430 730]{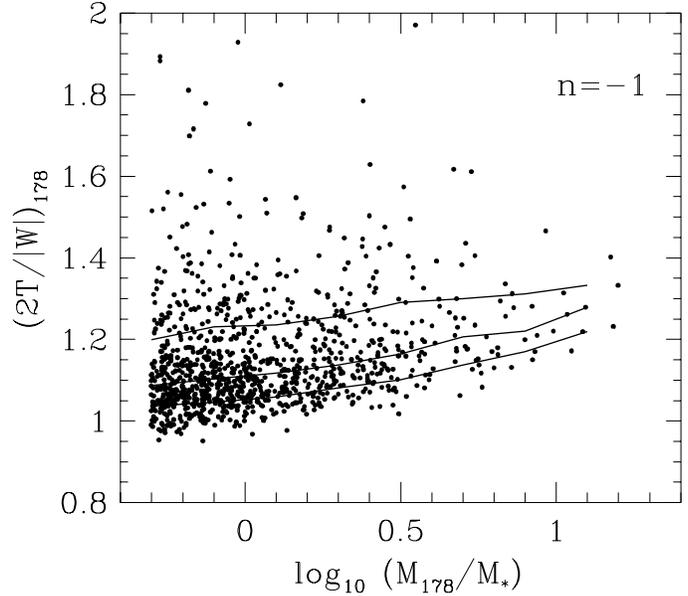}}
\caption{The scatter in the virial ratio evaluated interior to $\r178$
versus mass for the $n=-1$ simulation. 
The curves mark the locus of the 20th, 50th and 80th centiles 
of the distribution. 
The mass $\mstar$ equals 447 particles. }
\label{fig:scat_2tw}
\end{figure}

The distribution with mass of the ratio $(2T/|W|)_{178}$, evaluated in
spheres of radius $\r178$, is shown in Fig.~\ref{fig:scat_2tw}, for
the $n=-1$ simulation.  There is no tendency for the lower mass, less
well resolved haloes, to be further from virial equilibrium than the
larger mass haloes. In fact, the median value of $(2T/|W|)_{178}$ is
1.1 at low masses and increases gradually to approximately 1.2 at
$\m178/\mstar=10$.  This trend is completely consistent with the
variation of the halo density profiles reported above. The ratio
$2T/|W|$ at the virial radius can be computed for the \nfw model. For
$\an=0.1$, which fits the density profiles of haloes of mass $\m178
\approx \mstar$, $(2T/|W|)_{178}=1.13$, while for $\an=0.2$ which
corresponds to $\m178 \approx 10 \mstar$, $(2T/|W|)_{178}=1.23$. At
all masses there is a tail of objects with larger values of
$(2T/|W|)_{178}$. These are probably examples of ongoing mergers such
as depicted in Fig.~\ref{fig:dots}c.


\section{Angular Momentum}\label{sec:lambdas}

\begin{figure}
\centering
\centerline{\epsfxsize = 13cm \epsfbox[0 70 574
740]{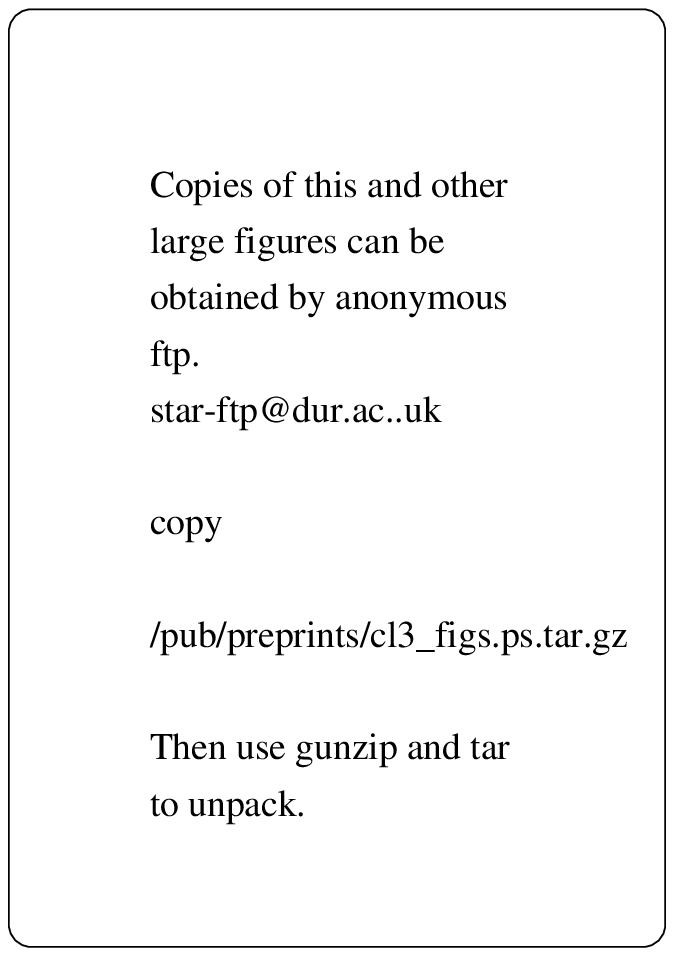}}
\caption{The distribution of the spin parameter $\lambda_{178}$,
evaluated for material within a sphere of radius $\r178$, against
mass. The lines mark the locus of the 20th, 50th and 80th centiles
of the distribution. 
Note that $\mstar$ equals $266$, $447$ and~$46.8$ in the $n=0$, $-1$ 
and~$-2$ simulations respectively.
}
\label{fig:scat_lambdas}
\end{figure}

\begin{figure}
\centering
\centerline{\epsfxsize = 15cm \epsfbox[0 70 574
720]{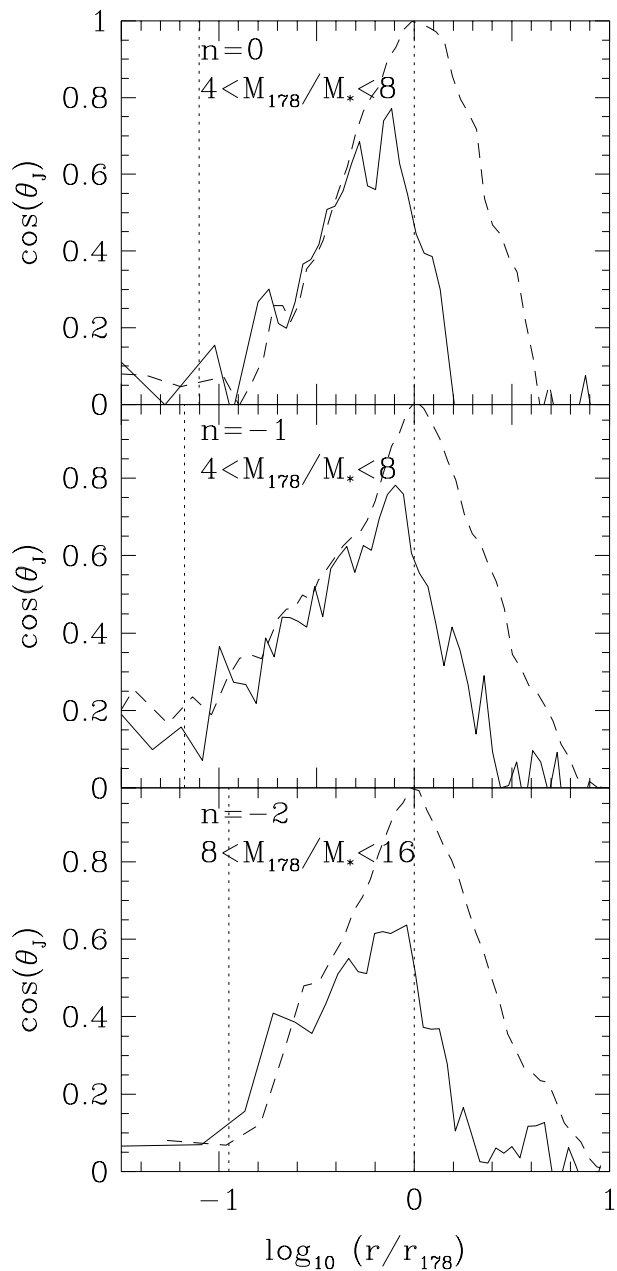}}
\caption{The alignment of the angular momentum within haloes.  The
angle $\theta_J$ is the angle between the total halo angular momentum,
computed from all the material interior to the virial radius $\r178$,
and the angular momentum of material in shells or spheres. The curves
give the mean value of $\cos(\theta_J)$ for shells (solid curves) and
spheres (dashed curves) where the averaging is done over all the
haloes in the specified mass ranges.  }
\label{fig:costhetaj}
\end{figure}

The dimensionless spin parameter
\begin{equation} 
\lambda = {{J \vert E \vert^{1/2} }\over { G M^{5/2}}},
\end{equation}
where $J$, $E$ and $M$ are the total angular momentum, energy and mass
is an important dynamical parameter that has been much studied in
numerical simulations ( e.g. Barnes \& Efstathiou 1987).  Our results
are in broad agreement with earlier work and have a median $\lambda
\approx 0.04$ at $M=\mstar$.  Fig.~\ref{fig:scat_lambdas} shows, as a
function of mass, the distributions of $\lambda_{\rm 178}$ evaluated
for spheres of radius $\r178$.  The median, 20th and 80th centiles are
indicated by the solid lines.  They reveal no obvious variation in the
distributions with spectral index, but a weak trend towards decreasing
values of $\lambda_{\rm 178}$ with increasing mass.  The earlier
outputs from the simulations confirm that this trend is real, and not
simply a result of the effective numerical resolution varying with
mass. The overall distributions and trends with mass are in good
agreement with those of Barnes \& Efstathiou (1987), who studied the
distribution of $\lambda$ for CDM and $n=0$ models, and those reported
by Efstathiou \etal (1988), who investigated other scale free
models. Efstathiou \etal  also found a weak dependence of the
median value of $\lambda$ on spectral index, but for a fixed
$\m178/\mstar$ we find any such dependence to be extremely weak. For
example, for $1/2<\m178/\mstar<2$, we find median values for
$\lambda_{\rm 178}$ of $0.045$, $0.043$ and $0.044$ for $n=0$, $-1$
and $-2$ respectively, with uncertainties of about $0.001$. For
$2<\m178/\mstar<4$, we find medians $0.034$, $0.037$ and $0.038$ with
uncertainties of around $0.002$. Efstathiou \etal may have found a
stronger trend because by averaging over all the haloes in each
simulation, they gave different weights to different ranges of
$M/\mstar$ for different $n$.

The degree of alignment of the angular momentum throughout each halo
is studied in Fig.~\ref{fig:costhetaj}. For each halo we measured the
angle $\theta_J$ between the angular momentum vector of material in
shells or spheres of radius $r$ and the total angular momentum vector
of the material within the virial radius $\r178$. As a function of
radius, Fig.~\ref{fig:costhetaj} shows the mean value of
$\cos(\theta_J)$ for shells (solid curves) and spheres (dashed curves)
for haloes selected in various mass ranges from each of the three
simulations.  If the angular momentum were randomly orientated then
$\langle\cos(\theta_J)\rangle =0$.  We find some degree of alignment
through the virialized region of the haloes, with a sharp drop in the
alignment beyond $\r178$. The estimate of the degree of alignment is
sensitive to the noise in estimates of $J$.  Thus, the true alignment
may be somewhat better than indicated by this analysis.  A more
detailed study of the angular momentum distribution can be found in
Warren \etal (1992).

\vfill
\eject

\section{Asymmetry and Substructure}\label{sec:asymm}

\begin{figure*}
\centering
\centerline{a) \epsfysize = 19.5cm \epsfbox{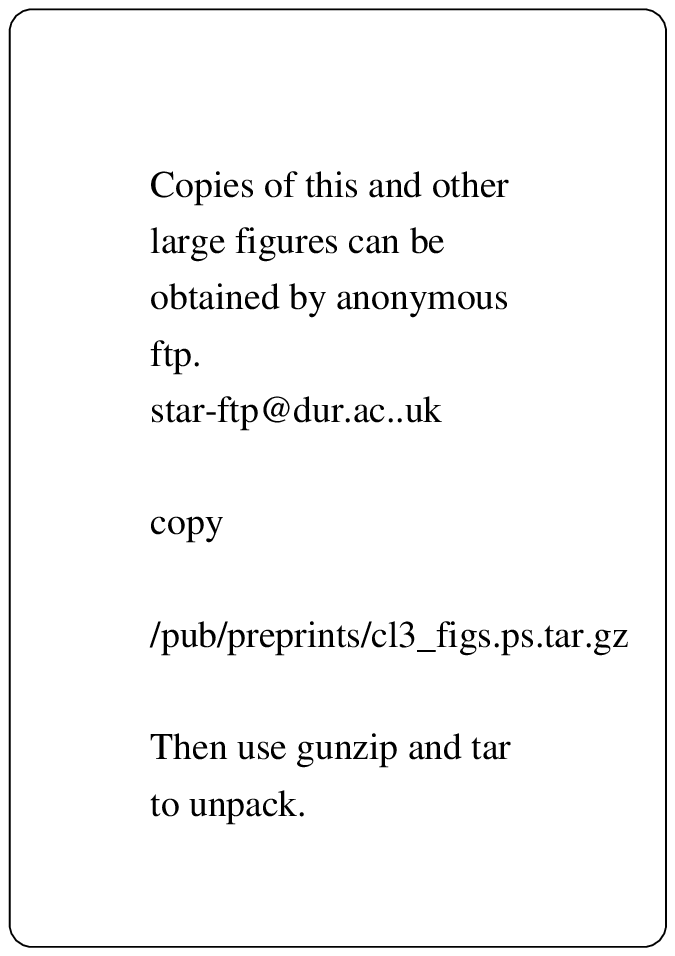} \quad \quad
\quad b) \epsfysize = 19.5cm \epsfbox{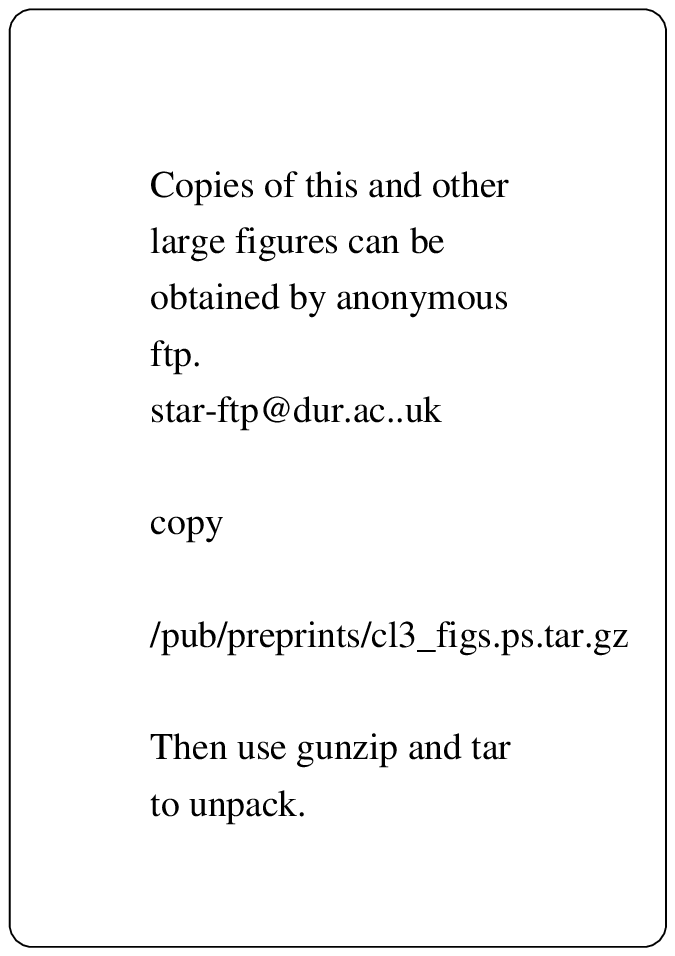} }
\caption{A selection of 4 groups from the $n=-1$ simulation  with
masses in the range $8<\m178/M_*<16$
($3576<N<7152$), showing the variety of group
morphologies and comparing the FOF(0.2) group definition with that of
SO(178). The particles plotted are those defined as a group by the
SO(178) algorithm.  Each group is shown projected along the three
principal axes defined by the inertia tensor of the material within
the sphere of radius $\r178$. In each panel the dashed circle marks
the projected boundary of this sphere.  The flattening and orientation
of the inertia tensor is indicated by the dotted ellipses.  The
contours show the projected density of all particles within
$2\r178$. The contour levels are $0.5$, $2$ and~$8$ times the mean
projected surface density within the $\r178$ circle.  The outer
connected contour corresponds quite accurately to the boundary of
the group defined by the FOF(0.2) algorithm.  The values of $d_{\rm
off}$ given in the upper panel of each figure are the offsets of the 
potential centre from the group centre of mass for the SO(178) and
FOF(0.2) group definitions.  }
\label{fig:dots}
\end{figure*}

\begin{figure*}
\centering
\centerline{c) \epsfysize = 19.5cm \epsfbox{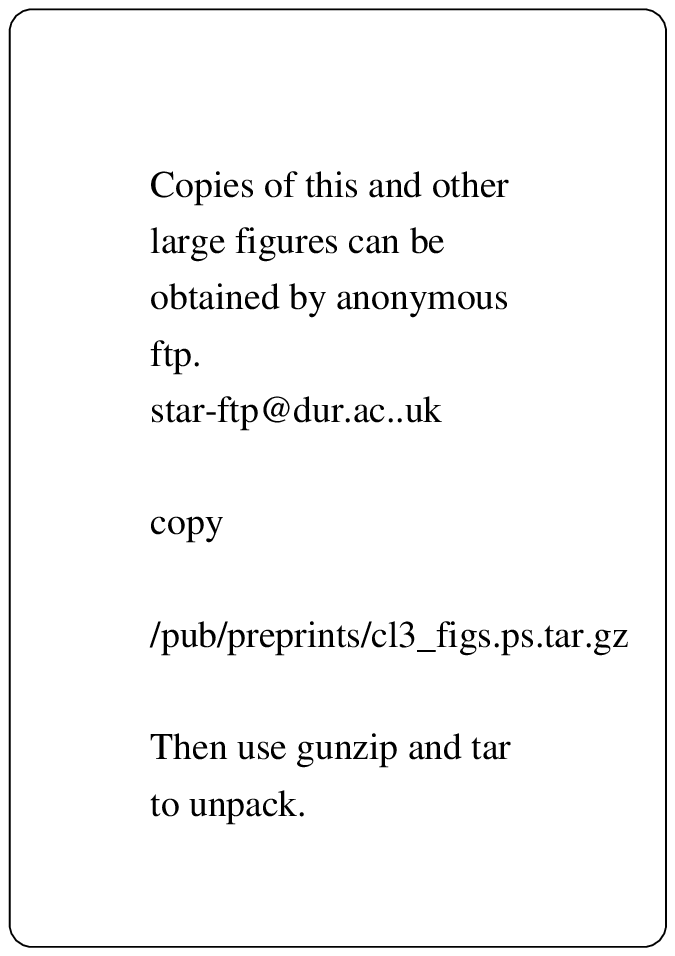}
\quad \quad \quad d) \epsfysize = 19.5cm \epsfbox{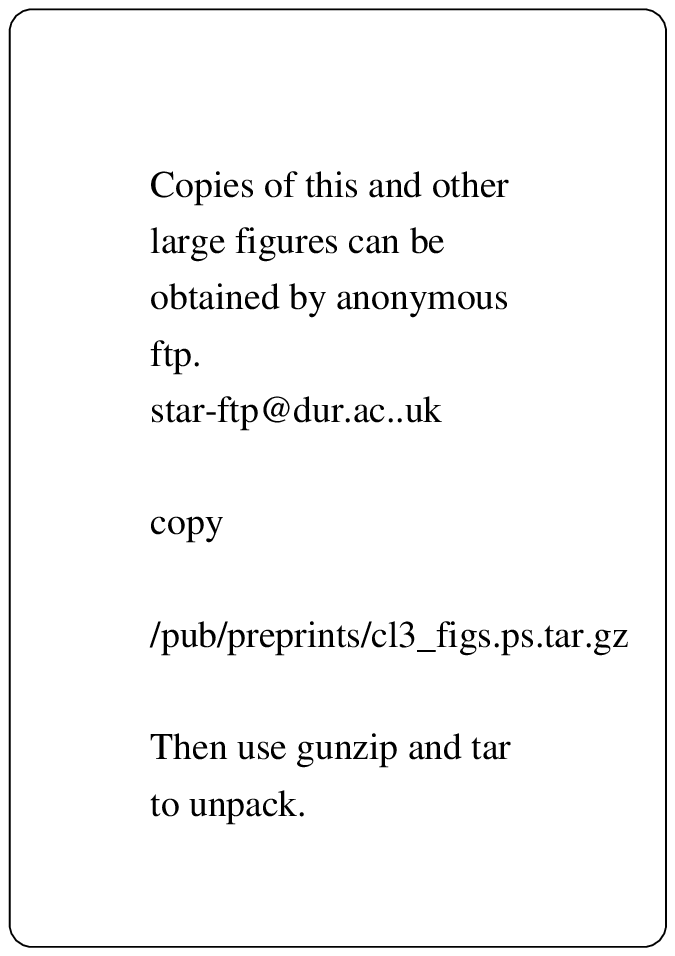}}
\leftline{{\bf Figure \ref{fig:dots}. } continued.}
\end{figure*}

\begin{figure*}
\centering
\centerline{\epsfbox[18 320 592 700]{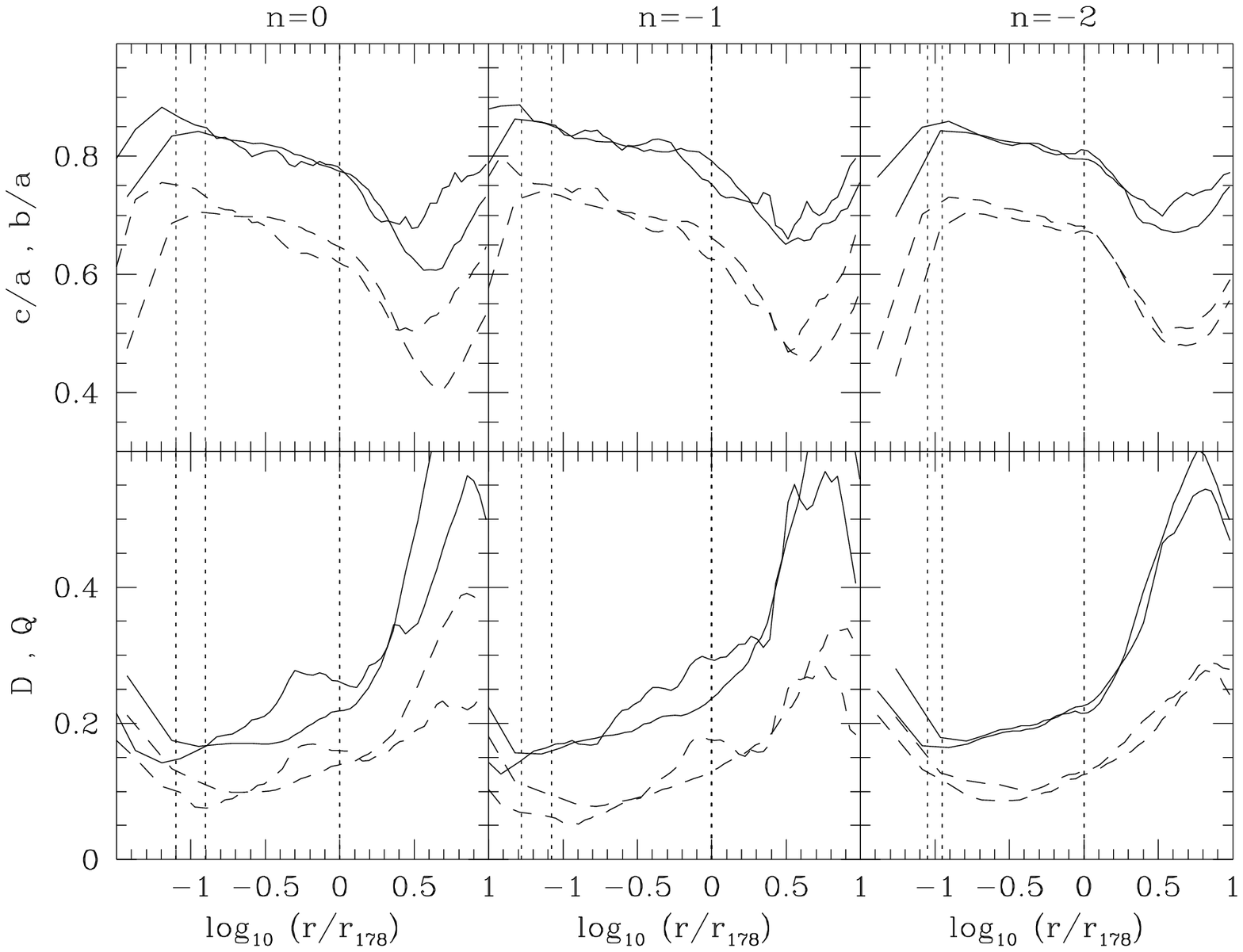}}
\caption{For each of the three simulations the curves in the upper
panels show the axial ratios $b/a$ (solid) and $c/a$ (dashed) 
of the inertia tensor of material in spheres.
The curves shown for $n=0$ are for masses $1<\m178/\mstar<2$ and
$4<\m178/\mstar<8$. For $n=-1$ curves for the two mass ranges
$2<\m178/\mstar<4$ and $8<\m178/\mstar<16$ are
shown. For $n=-2$ the mass ranges are $8<\m178/\mstar<16$ and
$32<\m178/\mstar<64$.  The curves in the lower panels show the
corresponding dipole, $D(r)$ (dashed), and quadrupole $Q(r)$ (solid)
values.  As before, the vertical dotted lines mark the
force-softening scale $\eta$ for the two mass ranges 
and the virial radius $\r178$.  
Note that $\mstar$ equals $266$, $447$ and~$46.8$ in the $n=0$, $-1$ 
and~$-2$ simulations respectively.
}
\label{fig:abc}
\end{figure*}

In the previous analysis, by spherically averaging we have treated the
haloes as if they were both smooth and intrinsically spherical.  In
reality, the haloes are non-spherical and contain significant
sub-structure. Figs.~\ref{fig:dots}a-d show examples of four groups
selected from the $n=-1$ simulation to illustrate the variety of
morphologies.  Each group is shown projected along each of the three
principal axes of the inertia tensor of the material within the radius
$\r178$, whose orientation is indicated by the dotted ellipse in each
panel. The particles plotted are those identified as group members by
the SO(178) algorithm. The contours show the projected surface
density of all particles within $2\r178$ of the group centre. For
these groups the outermost contours agree extremely well with the
projected boundaries of the same groups when identified using the
FOF(0.2) algorithm.  

The first two groups, Figs.~\ref{fig:dots}a-b,
are typical of the majority of groups.  There is a good correspondence
between the FOF(0.2) and SO(178) group definitions and both have
centres of mass that coincide quite well with the potential
centre. The offset between these two centres is indicated in the upper
panel of each figure.  The remaining two groups are examples where the
FOF(0.2) algorithm links together two distinct haloes which are in the
process of merging.  In such cases the offset between the potential
centre and group centre of mass can be large for the FOF(0.2) group,
but remains small for the SO(178) definition. Thus for the FOF(0.2)
groups, there is a correlation between the morphology and the offset
between the centre of mass and the potential centre. Approximately
20\% of FOF(0.2) groups have $d_{\rm off}>0.3 \r178$.

To quantify the distribution of halo shapes and their radial dependence
within individual groups, we have studied the first and second moments
of the mass distribution within spheres.

The first moments define a dipole vector, whose first component is
\begin{equation} 
L_x(r) = \sum_{\rm sphere} x/N_{\rm sphere},
\end{equation} 
where the summation is over all $N_{\rm sphere}$ particles interior to
radius $r$.  The statistic that we study is a normalised magnitude of
this vector, $D(r) = \vert {\bf L}(r) \vert/\langle r \rangle_r$,
where $\langle r \rangle_r$ is the mass weighted mean value of $r$
within the sphere. This definition implies that an isotropic
distribution has $D(r)=0$, while the maximum value of $D(r)$ is unity
corresponding to all the mass concentrated in one direction.

The second moments of the mass distribution define the components of
the inertia tensor. For example
\begin{equation} 
I_{xy} (r) = \sum_{\rm sphere} x y/N_{\rm sphere},
\end{equation} 
where the summation is again over all particles
interior to radius $r$.  If we denote the ordered eigenvalues of this
tensor as $a^2$, $b^2$ and~$c^2$, then the mass distribution can be
characterised by an ellipsoid with axial ratios $a:b:c$.  We study
these axial ratios as a function of radius $r$, and also a normalised
quadrupole statistic which may be written in terms of these
eigenvalues as
\begin{equation} Q(r) = \frac{\left(
2(a^2+b^2+c^2)^2-6(a^2b^2+b^2c^2+c^2a^2)\right)^{1/2}} 
{ \langle r^2 \rangle_r},
\end{equation} 
where $\langle r^2 \rangle_r$ is the mass weighted mean value of $r^2$ 
within the sphere.
An isotropic distribution has $Q(r)=0$ while the maximum
value is $Q(r)=\sqrt{2}$.  


Figure~\ref{fig:abc} shows the mean axial ratios and quadrupole and
dipole statistics for a selection of well resolved haloes from each
simulation. For fixed $\m178/\mstar$ poorly resolved haloes from the
earlier output time of our simulations have noisier and larger
quadrupoles. For all mass ranges the halo shapes become increasingly
aspherical -- small values of $b/a$ and $c/a$ and large values of
$Q(r)$ and $D(r)$ -- at radii smaller than the force
softening scale $\eta$. At larger radii, for haloes containing more
than approximately 100 particles within $\r178$, we find that these
profiles have very little systematic dependence on mass. One can also
see from Fig.~\ref{fig:abc} that the profiles also have negligible
dependence on the spectral slope of the initial conditions.  The
typical mean axial ratios $a:b:c$ are $1:0.8:0.65$ at the virial
radius $\r178$.  The axial ratio profiles indicate that the haloes
gradually become more spherical with decreasing radius until one
approaches the force softening scale. The dipole and quadrupole
statistics are slowly increasing throughout the range $\eta<r<\r178$,
with typical values of $0.15$ and $0.25$ at the virial radius.  Beyond
the virial radius, the random distribution of the material falling
towards the halo results in a sharp decline in the axial ratios $b/a$
and $c/a$ and a corresponding sharp increase in both $D(r)$ and
$Q(r)$.

\begin{figure}
\centering
\centerline{\epsfxsize = 16cm \epsfbox[0 80 574 715]{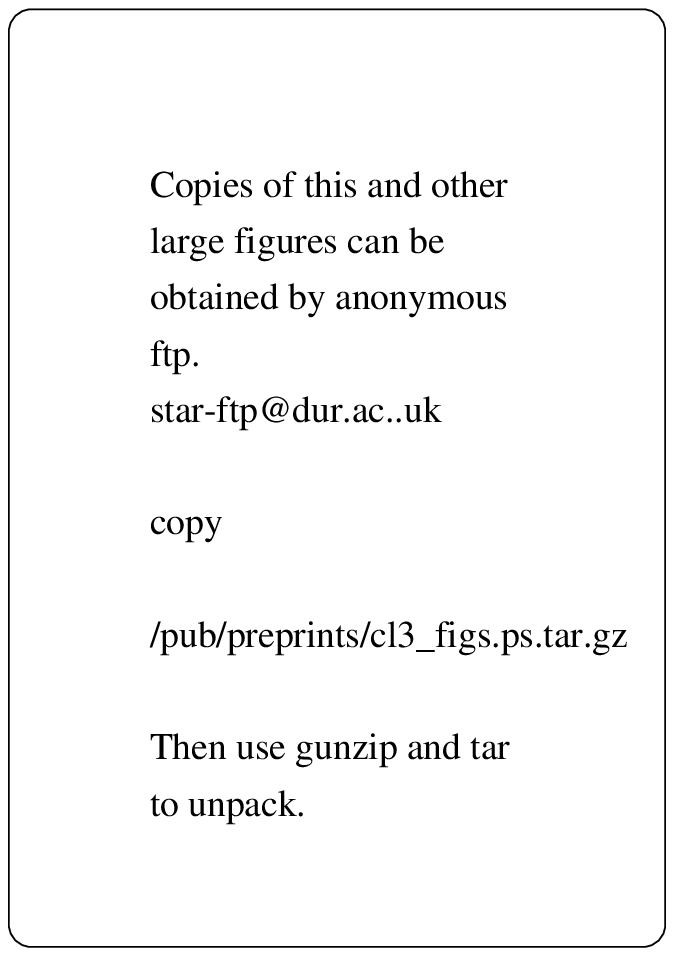}}
\caption{The distribution of the axial ratios $b/a$ and $c/b$ for the
interia tensor of the mass interior to $\r178$ for SO(178) groups with
masses $\m178\ge\mstar$ from each of the three simulations.  The locii
of prolate ($c/b=1$) and oblate ($b/a=1$) shapes are indicated.  }
\label{fig:shapes}
\end{figure}

  The distribution of axial ratios $a:b:c$ at the virial radius is
shown in Fig.\ref{fig:shapes}.  This figure combines all haloes with
masses $\m178>\mstar$, but splitting the samples reveals no
significant mass dependence.  For each spectral index the distribution
of axial ratios is broad, with all haloes being to some degree 
triaxial. For spectral indices $n=-1$ and~$0$ prolate configurations
($c/b < b/a$) are mildly favoured over oblate ($c/b > b/a$), but no
such trend is evident in the case of $n=-2$.

Our conclusions about axial ratios are in general agreement with those
of Efstathiou \etal (1988). Our values for the typical axial ratios
also agree with Warren \etal (1992), when for the latter we compare to
the values they calculate from the inertia tensor within
spheres. Warren \etal also use a method of fitting ellipsoids to the
density distribution, from which they find a trend for haloes to
become more spherical with increasing radius, opposite to the trend we
find. 

\vfill


\section{Discussion}\label{sec:discuss}

To the limit of the resolution of our simulations, the densities of
dark matter haloes diverge as one approaches their centres. As a first
approximation, the density profiles between the gravitational
softening scale and the virial radius are fit by power laws, in
agreement with Crone \etal (1994). For $\Omega=1$ and $n=-2$, the
power law is close to $\rho\propto r^{-2}$, consistent with flat
rotation curves, but the effective power-law slope steepens
significantly with increasing $n$ and with decreasing mass.  However,
the density profiles have significant curvature in the $\log\rho$ --
$\log r$ plane, steepening with increasing radius. Better 2-parameter
fits than a power law $\rho\propto r^{-\alpha}$ to the spherically
averaged density and circular velocity profiles are provided by the
Hernquist \shortcite{hq} or Navarro \etal (1995a,b) analytical models,
which each have a scale radius $b$ , and thus a logarithmic slope
which changes with radius. Both of these models have density profiles
which are shallower than $r^{-2}$ at $r\ll b$ and steeper at $r\gg
b$. The Hernquist model changes slope too abruptly to match the
simulated haloes over the full range of resolved scales, but the \nfw
model is an excellent fit to all our mean halo profiles.  The
\nfw model also reproduces other dynamical properties of the simulated
haloes interior to their virial radii $\r178$.  Ignoring the small
radial anisotropy of the halo velocity dispersion tensor, the slow
variation of velocity dispersion with radius is well matched by that
in the isotropic \nfw model. Once the radius $\r178$ has been measured
for a halo, the \nfw model has only one free parameter, a
dimensionless scale radius $\an=\bn/\r178$. It is remarkable that such
a simple one-parameter function is able to fit the density profiles of
haloes over a wide range of masses and initial conditions. Formally,
the Hernquist and \nfw models imply density varying as $\rho\propto
r^{-1}$ at very small radii ($r\ll b$), but even in our best resolved
haloes, our profiles only extend down to $r/\bn \approx 0.3$, where
the slope for the NFW model is approximately $r^{-1.5}$, so we have
not proved that the asymptotic $r^{-1}$ dependence is correct.

The dependence of the dimensionless scale radius, $a$, on halo mass
and on the slope of the initial power spectrum is of particular
interest.  Lacey \& Cole (1993,1994) demonstrated that higher mass
haloes have higher accretion rates and more recent formation epochs
than their lower mass counterparts, where the formation time is
defined as the time at which half the present mass of a given halo had
already been assembled. Also, for a given $\m178/\mstar$.  the halo
formation time is earlier for increasing spectral index $n$. If the
inner regions or ``core'' of the final halo formed mostly from the
material which assembled earlier, then they should reflect the
background density of the universe at the time they were assembled,
and be denser for haloes which formed earlier. Therefore one predicts
denser, more concentrated, cores for haloes with earlier formation
times.  This is precisely the trend we find in the simulated haloes.
For fixed $n$, the lower mass haloes are more concentrated, with
smaller values of $a$, while at fixed mass, $a$ decreases with
increasing $n$.  Navarro \etal (1995b) investigate the first of these
trends for the CDM model, and find a quite good proportionality
between the core density and the density of the universe at the
formation time.

The dependence of the density profiles and rotation curves on the
spectral index is also in qualitative agreement with the results of
Quinn \etal (1986), Efstathiou \etal (1988) and Crone \etal (1994),
who found steeper profiles for power spectra with more small scale
power, \ie increasing $n$.   Crone \etal (1994) found no significant
dependence of the density profiles on halo mass.  However, they
examined a smaller range of halo masses than in the present
paper. They also applied a correction procedure intended to compensate
for the effect of the finite force resolution in their
simulations. This correction made all their haloes have steeper core
density profiles. We tested this procedure on our haloes with similar
results. However, we did not find that it improved the agreement
between the halo profiles with the same $\m178/\mstar$ from two
different output times. Thus, although the procedure alters the
density profiles, it is not clear that it is usefully correcting for
all the effects of limited resolution, which include not only force
resolution but also finite particle number and limited range of scales
in the initial conditions. Thus we preferred to leave the density
profiles uncorrected and instead fit the analytical models only over
the range $\eta<r<\r178$, where comparison of the different output
times showed the density profiles to be robust.

We can also compare our results with the analytical model of secondary
infall around density peaks developed from the work of Gunn \& Gott
(1972) by Hoffman \& Shaham (1985). Hoffman \& Shaham predicted
power-law halo density profiles, with logarithmic slope equal to $-2$
for $n\leq -1$ and $-(9+3n)/(4+n)$ for $n>-1$. We find that the
profiles depart from power laws, though the average profile slopes are
similar to the Hoffman-Shaham values over the regions where we can
measure them, and Hoffman-Shaham predicts the correct trend of steeper
profiles with increasing $n$. However, the Hoffman-Shaham model also
predicts that the profile shapes should be independent of halo mass,
which disagrees with our N-body results.

We have investigated how best to draw the boundary between the region
of a halo which is virialized and in approximate dynamical equlibrium,
and the outer parts which are still infalling. This is particularly
important if one wants to define a total virialized mass for the halo,
to use in comparing with theoretical models which express results in
terms of a total halo mass. An example of the latter is the
Press-Schechter model for the mass function of haloes (Press \&
Schechter 1974), and its extension to statistics of halo mergers
(Lacey \& Cole 1993,1994). Our starting point was the spherical
collapse model, which for the collapse of a uniform sphere in an
$\Omega=1$ universe, predicts that the virialized region lies interior
to a radius $\r178$ for which the mean density is $178$ times the
background value. We then find that the mean radial velocity profiles
plotted in the dimensionless form $V_r/\v178$ {\it vs.} $r/\r178$ all
look very similar for different $n$ and different $\m178/\mstar$, and
all show a break at $r\approx\r178$ from a quasi-static interior with
$V_r\approx 0$ to an infalling exterior (for $1\lta r/\r178 \lta 4$)
with $V_r<0$. Thus, the radius $\r178$, and the corresponding enclosed
mass $\m178$, do indeed provide good characterizations of the radius
and mass of the virialized region. The shape of the mass distribution
also becomes significantly more non-spherical beyond $\r178$,
consistent with a break between dynamically relaxed inner regions and
unrelaxed outer regions, and supporting the choice of $\r178$ as the
virial radius. On the other hand, the virial ratio $2T/|W|$ in spheres
varies smoothly through $\r178$, and at $\r178$ exceeds the value
$2T/|W|=1$ for an isolated object in dynamical equilibrium by
10--20\%, principally because of the confining effect of the external
infalling material. Thus, the value of $2T/|W|$ does not provide a
good criterion for deciding where the virialized region ends.

Crone \etal (1994) followed a similar procedure to us, but instead
chose an overdensity of 300 to define the edge of the virialized
region. In practical terms, the difference from our choice is not that
large. For the haloes in our simulations, the dimensionless scale
radius spans the range $0.05\lta \an \lta 0.3$. For this range, the
\nfw model gives  $r_{300}/\r178\approx 0.8$ and $M_{300}/\m178
\approx 0.8-0.9$, where $r_{300}$ and $M_{300}$ are defined
analogously to $\r178$ and $\m178$. 

The halo properties we analysed are largely independent of the
algorithm used to identify the groups, for the SO method with mean
overdensity in the range $\kappa=100-400$, and for the FOF method with
linking length in the range $b=0.15-0.3$, corresponding to a local
overdensity in the range $20-140$.  The halo profiles are insensitive
to exactly which particles are identified as group members because we
use the minimum in the gravitational potential to define the halo
centre, and include all the surrounding material when constructing the
halo profiles.

What group-finder partitions the particle distribution into objects
that best correspond to virialized haloes according to the results
above? Not surprisingly, the raw groups found by the SO(178) algorithm
mostly match very well with the material interior to $\r178$ around
the halo centre, defined as the minimum in the gravitational
potential. The difference is that the raw SO(178) groups are defined
so that the centre of mass of the group coincides with the centre of
the sphere. The FOF(0.2) groups mostly also match fairly well to the
material within $\r178$ of the halo centre, as defined above, but
occasionally the FOF(0.2) algorithm links two distinct haloes which
are soon to merge, and in such cases the offset between the FOF(0.2)
group centre of mass and the potential centre can be large. The masses
of the FOF(0.2) groups mostly agree well with the virial masses
$\m178$, with a spread around this of only about 20\%, but with larger
deviations for haloes in the process of merging. This provides some
{\it post hoc} justification for the extensive use of the FOF(0.2)
algorithm in computing halo properties and mass functions.

\section{Conclusions}\label{sec:conc}

We have studied the structure of haloes that are formed in simulations
of self-similar clustering models with $n=0,-1,-2$ and $\Omega=1$,
over a wide range in halo mass.

(1) We find that the radius $\r178$ about the halo centre within which
the mean overdensity is 178 accurately demarcates the virialized 
interior of the halo, which is in approximate dynamical equlibrium,
from the exterior, where material is still falling in. The
characterization of $\r178$ as the ``virial radius'' agrees with the
prediction of the simple spherical collapse model.

(2) For $r<\r178$, the spherically averaged density, circular velocity
and velocity dispersion profiles are very well fit by the analytical
model of Navarro \etal (1995a,b) with an isotropic velocity
dispersion. This model has $\rho \propto r^{-1}$ at $r/\r178\ll \an$,
rolling over to $\rho \propto r^{-3}$ at $r/\r178\gg \an$, where $\an$
is a dimensionless scale radius. It should noted that even for out best 
resolved haloes the \nfw model has $\rho \propto r^{-1.5}$ at 
the softening radius, and thus we do not probe the asymptotic prediction of 
$\rho \propto r^{-1}$. 

(3) The value of the dimensionless scale radius $\an$ correlates
strongly with halo mass and with spectral index $n$. Haloes with mass
$M$ much less than the characteristic mass $\mstar$ typically form
much earlier than haloes with $M\gg\mstar$. This results in them being
more centrally concentrated (smaller $\an$), with cores whose
densities reflect the higher mean density of the universe at their
formation time. For a given $M/\mstar$, haloes form earlier in models
with more small-scale power (larger $n$), and also have denser cores.

Despite the successes of the spherically symmetric model, haloes are
not spherical.  In these $\Omega=1$ simulations most haloes show
evidence of substructure and many are clearly about to undergo a
merger. 

(4) Haloes are generically triaxial, but with a slight preference for
prolate configurations, at least for $n=-1$ and~$0$. Their inertia
tensors (measured in spheres) have typical axial ratios $1:0.8:0.65$
at the virial radius $\r178$, becoming gradually more spherical
towards their centres. There is no significant trend with mass or
spectral index. Measurements of the dipole moments of the mass
distribution around  halo centres show them to have mean dipoles of
order 10\%, and similarly large quadrupole distortions.

(5) Haloes are slowly rotating, with $V_{rot}/\sigma\approx 0.3$. The
rotation velocity $V_{rot}$ measured in spherical shells is
approximately constant with radius, but the angular momentum is not
well aligned between the central and outer regions.  The median value
of the spin parameter (measured interior to $\r178$) is $\lambda
\approx 0.04$ with a weak trend for lower $\lambda$ at higher halo
mass. There is no significant trend of $\lambda$ with spectral index.

(6) The properties we find are insensitive to the details of the
group-finding algorithm used to select the haloes in the simulations,
since once we have identified a group, we define the halo centre from
the minimum of the potential, and include all surrounding particles
when computing profiles.

The simple analytical description in (1)--(3) of the spherically
averaged profiles of dark matter haloes has many applications. It will
enable refined modelling of galaxy formation and dynamics, and improved
calculations of gravitational lensing properties.  A similarly
detailed analysis of haloes formed in open models and models with a
cosmological constant will be extremely interesting, and may lead to
new diagnostics in which the halo density profiles can be used to
discriminate between different cosmological models.

\section*{ACKNOWLEDGEMENTS}

We thank George Efstathiou for supplying us with a copy of his P$^3$M
N-body code. SMC gratefully acknowledges the support of a PPARC
Advanced Fellowship. CGL was supported by a PPARC Advanced Fellowship
at Oxford, and by the Danmarks Grundforskningsfond through its support
for the establishment of the Theoretical Astrophysics Center.


\begin{thebibliography}{}
\bibitem[\protect\citename{Barnes \& Efstathiou }1987]{be}
Barnes, J., Efstathiou, G. P. E., 1987 ApJ 319,575 
\bibitem[\protect\citename{Cole \etal }1994]{cafnz}
Cole S., Aragon-Salamanca A., Frenk C. S., Navarro J. F., Zepf, S., 1994 
MNRAS, 271,781
\bibitem[\protect\citename{Crone \etal }1994]{crone}
Crone, M. M., Evrard, A. E., Richstone, D. O., 1994 ApJ 434,404
\bibitem[\protect\citename{Davis \etal }1985]{defw}
Davis M., Efstathiou G., Frenk C. S., White, S. D. M., 1985,
ApJ, 292, 371
\bibitem[\protect\citename{Dubinski \& Carlberg }1991]{dub}
Dubinski, J., Carlberg, R., 1991 ApJ 378,496
\bibitem[\protect\citename{Efstathiou \etal }1985]{edfw}
Efstathiou G., Davis M., Frenk, C. S., White, S. D. M., 1985,
ApJ Supp, 57, 241 
\bibitem[\protect\citename{Efstathiou \etal }1985]{efwd}
Efstathiou G., Frenk, C. S., White, S. D. M., Davis M.,  1988,
MNRAS, 235,715
\bibitem[\protect\citename{Frenk \etal }1988]{fwde}
Frenk, C. S., White, S. D. M., Efstathiou, G., Davis, M., 1985 Nature 317,595
\bibitem[\protect\citename{Frenk \etal }1988]{fwde}
Frenk, C. S., White, S. D. M., Davis, M., Efstathiou, G., 1988 ApJ 327,507
\bibitem[\protect\citename{Gunn \& Gott }1972]{}
Gunn, J. E., Gott, J. R., 1972, ApJ, 176,1
\bibitem[\protect\citename{Hernquist }1990]{hq}
Hernquist L.,  1990, ApJ, 356, 359
\bibitem[\protect\citename{Hoffman \& Shaham }1985]{}
Hoffman, Y., Shaham J., 1985 ApJ, 297, 16
\bibitem[\protect\citename{Lacey \& Cole }1993]{lc1}
Lacey C. G.,  Cole S., 1993, MNRAS, 262,627
\bibitem[\protect\citename{Lacey \& Cole }1994]{lc2}
Lacey C. G.,  Cole S., 1994, MNRAS, 271,676 
\bibitem[\protect\citename{Kauffmann, \etal }1993]{kwg}
Kauffmann G., White S. D. M., Guiderdoni B., 1993, MNRAS 264, 201
\bibitem[\protect\citename{Narayan \& White }1988]{nw}
Narayan R., White S. D. M., 1988, MNRAS, 231, 97P
\bibitem[\protect\citename{Navarro \etal }1995a]{nfwa}
Navarro, J. F., Frenk, C. S., White, S. D. M., 1995a, MNRAS 275,720 NFW
\bibitem[\protect\citename{Navarro \etal }1995b]{nfwb}
Navarro, J. F., Frenk, C. S., White, S. D. M., 1995b, MNRAS, in press.
\bibitem[\protect\citename{Press \& Schechter }1974]{ps}
Press, W. H., Schechter P., 1974, ApJ 187, 425
\bibitem[\protect\citename{Quinn \etal }1986]{qsz}
Quinn P. J., Salmon, J. K., Zurek W. H., 1986 Nature 322, 329
\bibitem[\protect\citename{Warren \etal }1992]{warren}
Warren, M. S., Quinn, P. J., Salmon, J. K., Zurek, W. H., 1992 ApJ 399,405
\bibitem[\protect\citename{Zurek \etal }1988]{zqs}
Zurek, W. H., Quinn, P. J., \& Salmon, J. K., 1988 ApJ 330, 519
\end{thebibliography}
\end{document}